\documentclass[aps,pre,twocolumn,nofootinbib]{revtex4-1}
\usepackage[dvips]{graphicx,rotating}
\usepackage{amssymb}
\usepackage{amsmath}
\usepackage{epsfig}
\usepackage{bm}
\usepackage{amsfonts,float}
\usepackage{hyperref}

\usepackage{graphicx}
\usepackage[usenames,dvipsnames]{color}

\begin{document}

\title{Ergodicity and large deviations in physical systems with stochastic dynamics}

\author{Robert L. Jack}
\affiliation{Department of Applied Mathematics and Theoretical Physics, University of Cambridge, Wilberforce Road, Cambridge CB3 0WA, United Kingdom}
\affiliation{Department of Chemistry, University of Cambridge, Lensfield Road, Cambridge CB2 1EW, UK}
\email{rlj22@cam.ac.uk}

\begin{abstract}
In ergodic physical systems, time-averaged quantities converge (for large times) to their ensemble-averaged values.  Large deviation theory describes rare events where these time averages differ significantly from the corresponding ensemble averages.  It allows estimation of the probabilities of these events, and their mechanisms. This theory has been applied to a range of physical systems, where it has yielded new insights into entropy production, current fluctuations, metastability, transport processes, and glassy behaviour.  We review some of these developments, identifying general principles.  We discuss a selection of dynamical phase transitions, and we highlight some connections between large-deviation theory and optimal control theory.
\end{abstract}

\newcommand{\beq}{\begin{equation}}
\newcommand{\eeq}{\end{equation}}
\newcommand{\eps}{\epsilon}
\newcommand{\mS}{m_{\rm S}}
\newcommand{\mU}{m_{\rm U}}
\newcommand{\ee}{{\rm e}}

\newcommand{\CC}{\mathcal{C}}
\newcommand{\Fbar}{\overline{F}}
\newcommand{\bbar}{\overline{b}}

\maketitle

\section{Introduction}
\label{sec:introduction}

In statistical mechanics, many properties of equilibrium systems can be calculated using free-energy methods, and the underlying Boltzmann distribution.
%
%
However, this approach has two important restrictions -- it only applies in equilibrium, and it is restricted to static properties.  For example, the Boltzmann distribution has very little to say about dynamical quantities like viscosity and thermal conductivity, nor can it predict the time required for a protein to fold.  Predicting such quantities requires some knowledge of the equations of motion of a system: the relevant statistical mechanical theories must include \emph{dynamical information}.  
Such theories are useful in many contexts, which include
non-equilibrium steady states~\cite{Groot1984,Bertini2015,Seifert2012review} as well as dynamical aspects of the equilibrium state (for example in glassy materials~\cite{Cavagna2009}).  Other physical phenomena also involve transient relaxation to equilibrium, for example nucleation~\cite{Auer2001} and self-assembly~\cite{Whitelam2015annrev}.

For complex systems (with many strongly-interacting components), dynamical theories often assume that the behaviour is \emph{ergodic}.  That is, the  systems have steady states in which  time-averaged measurements converge (for long times) to corresponding ensemble averages.  Many important physical systems have this property, which motivates several questions.  For example: (i) How long does it take for the time-averaged measurements to converge?  (ii) What is the probability that the time-average does not converge to the ensemble average, given some long time $\tau$?   

In systems with deterministic dynamics, there is a rich and complex mathematical structure that allows such questions to be addressed, 
but the resulting theory has many subtle features~\cite{Ruelle2004today,Gallavotti1995,Gaspard1995}.  
Here we focus on stochastic processes, where the situation is somewhat simpler.  
In particular, the mathematical theory of large deviations~\cite{denH-book} can be used to analyse time-averaged quantities, as demonstrated by 
 important work in the late 1990s and early 2000s~\cite{Eyink1996,Lebowitz1999,Derrida1998,Bertini2002,Bodineau2004,Derrida2007}.
The theory has been applied to a range of physical systems, where it has provided new insights.  Examples include exclusion processes~\cite{Bertini2015,Derrida1998,Bertini2002,Bodineau2004,Derrida2007}, glassy materials~\cite{Garrahan2007,Hedges2009,Speck2012-sens,Malins2012-sens,pinch17}, models of heat transport~\cite{Hurtado2009,Lecomte2010,Ray2019-prb}, proteins~\cite{Weber2014protein,Mey2014}, climate models~\cite{Ragone2018}, and non-equilibrium quantum systems~\cite{Garrahan2010}.

This article outlines the application of large deviation theory as it applies to time-averaged quantities, and it describes some of the results and insights that have been obtained for physical systems.  By considering a range of applications, the aim is to complement other papers that focus primarily on the general structure of the theory~\cite{Touchette2009} or on specific classes of system~\cite{Derrida2007,Bertini2015}.  The remainder of this Section lays out some general principles and describes the theoretical context in more detail.  Later Sections are devoted to general aspects of the theory and to application areas including phase transitions, glassy systems, entropy production, and exclusion processes.  A few examples are discussed in detail.  The choice of applications and examples is biased towards the author's own work; they are presented within the broader context of the field.

\subsection{Fluctuations of time-averaged quantities}
\label{sec:intro-fluct}

This section introduces the main question that will be considered below.
Consider a system with stochastic dynamics, whose configuration at time $t$ is $\CC_t$.  Define an observable quantity $b=b(\CC)$ and a time interval $[0,\tau]$; then the time-average of $b$ is
\beq
\bbar_\tau = \frac{1}{\tau} \int_0^\tau b(\CC_t) \mathrm{d}t \; .
\label{equ:bbar}
\eeq
As a simple example one may consider an Ising model with $N$ spins, as in~\cite{Jack2010,Loscar2011}.  
Then $\CC=(\sigma_1,\sigma_2,\dots,\sigma_N)$ where each spin $\sigma_i=\pm1$.  Take $b(\CC)$ to be the energy of this configuration, so $\bbar_\tau$ is the time-averaged energy.  Clearly $\bbar_\tau$ is a random variable: different trajectories of the system have different values for this quantity.  However, in ergodic systems the typical situation is that $\bbar_\tau$ obeys a central limit theorem at large times: its distribution is Gaussian with a variance that decays as $\tau^{-1}$.  Motivated by  questions (i) and (ii) above, this article considers fluctuations that are not covered by the central limit theorem: large deviation theory is used to characterise \emph{rare events} where $\bbar_\tau$ differs significantly from its mean value, even as $\tau\to\infty$.  We will see below that these are \emph{exponentially} rare, in the sense that their log-probability is negative and proportional to $\tau$.

Since these events are very rare, one might wonder what relevance they have for practical physical systems.  In response to this question, we make two general points, which will be clarified below.
First, large-deviation theory has a rich structure and enables sharp statements about the dynamical behaviour of complex systems.  As such, it can be viewed as an idealised theoretical starting point for studies of dynamical behaviour in non-equilibrium systems, which enables general insight.  An important example is the analysis of fluctuation theorems~\cite{Lebowitz1999}.  Second, the theory has already proven useful for understanding the behaviour of physical systems, for example through analysis of metastable states in glassy systems~\cite{Hedges2009,Jack2011} and biomolecules~\cite{Weber2014protein}, and through uncertainty bounds on fluctuations of the current~\cite{Gingrich2016}, which are relevant for rare events and for typical fluctuations.

\subsection{Theoretical context}

The mathematical theory for large deviations of time-averaged quantities in stochastic processes was formulated by Donsker and Varadhan in the 1970s~\cite{DonVarI,DonVarII,DonVarIII,DonVarIV}. A clear presentation of the general (mathematical) theory of large deviations is given in the book of den Hollander~\cite{denH-book}.  An alternative mathematical approach to these problems is discussed in the book of Dupuis and Ellis~\cite{dupuis-book}, including a connection to ideas of optimal control theory, as discussed below.  In physical studies of non-equilibrium systems, work by Derrida and Lebowitz~\cite{Derrida1998} and Lebowitz and Spohn~\cite{Lebowitz1999} laid the foundations for the work described here, building on earlier studies~\cite{Gwa1992,Eyink1996,Gaspard1998}.
As mentioned in the introduction, theories of ergodicity and time-averages in deterministic systems also have a long history~\cite{Ruelle2004today,Gallavotti1995,Gaspard1995,Eckmann1985}, and large deviation theory is also relevant in these cases~\cite{Gallavotti1995,Gaspard1995,Ruelle2004}.  This article is restricted to stochastic systems, analysis of deterministic systems requires a different set of methods and assumptions.

 A separate strand of mathematical work applied large deviation theory to hydrodynamic limits~\cite[Ch. 10]{kipnis-landim-book}, and underlies the macroscopic fluctuation theory of Bertini, de Sole, Gabrielli, Jona-Lasinio and Landim~\cite{Bertini2002,Bertini2015}, which can also be used to analyse fluctuations of time-averaged quantities.  
 Yet another direction is the connection between large deviation theory and the theory of equilibrium statistical mechanics, as discussed by Ellis~\cite{Ellis-book}, 
 see also~\cite{Ruelle2004,sokal93}.
 
A useful resource from the physics literature is the review of Touchette~\cite{Touchette2009} which gives a clear presentation of large-deviation theory as it applies to equilibrium statistical mechanics and to time-averaged quantities, see also~\cite{Lecomte2007,Derrida2007}.  Two recent papers by Ch\'etrite and Touchette~\cite{Chetrite2015,Chetrite2015var} provide a comprehensive summary of the large deviation theory of time-averaged quantities, as it applies to physical systems.

\subsection{Outline}

The remainder of this article is structured as follows.  Sec.~\ref{sec:theory} gives an overview of the large deviation theory for time-averaged quantities.  It focusses on finite systems, which simplifies the analysis.  Sec.~\ref{sec:phase-transition} discusses some of the dynamical phase transitions that can occur in infinite systems, including an example calculation for the $1d$ Glauber-Ising model and a discussion of dynamical phase coexistence.  In Sec.~\ref{sec:glass} we discuss the behaviour of glassy systems, including dynamical phase transitions in kinetically constrained models.  Sec.~\ref{sec:fluct} discusses the role of time-reversal symmetries and large deviations of the entropy production, including an example from active matter.  We give a short discussion of exclusion processes and hydrodynamic behaviour in Sec.~\ref{sec:exc} before ending in Sec.~\ref{sec:outlook} with an outlook and a discussion of some possible future directions.

\section{General Theory}
\label{sec:theory}

This section outlines the general theory of large deviations of time-averaged quantities.  
This presentation is not at all complete, the aim is to highlight useful facts, in order to provide physical insight and intuition.
Nevertheless, some mathematical precision is required, in order to understand the scope and applicability of the theory; some technical details are provided in footnotes.  A more comprehensive presentation of similar material is given by Ch\'etrite and Touchette~\cite{Chetrite2015,Chetrite2015var}.

\subsection{Definitions}
\label{sec:def}

The central quantities that appear in large deviation theory are probability distributions, rate functions, and cumulant generating functions.  
These are introduced in a general context, some of the systems to which the theory can be applied are discussed in Sec.~\ref{sec:scope} below.
We consider models that converge at long times to unique steady states, and angled brackets $\langle \cdot \rangle$ indicate steady-state averages.

Recalling (\ref{equ:bbar}),
the probability density for $\bbar_\tau$ is denoted by
$
p(\bbar|\tau) 
$
. 
The \emph{cumulant generating function}
(CGF) for $\bbar_\tau$ is
\beq
G(s,\tau) = \log \big\langle \ee^{-s \tau \bbar_\tau} \big\rangle \; .
\label{equ:cgf}
\eeq
One sees that $G(0,\tau)=0$ and $(\partial G/\partial s)_{s=0} = -\langle \tau \bbar_\tau \rangle$.\footnote{%
Our definitions mostly follow~\cite{Garrahan2009}, in particular
we include a minus sign in the exponent of (\ref{equ:cgf}), which is natural for the thermodynamic analogy discussed in Sec.~\ref{sec:thermo}.
However, analogous definitions of the CGF without any minus sign are also common in the literature.  
}

To analyse large deviations, we consider the limit of large $\tau$, defining
\beq
I(\bbar) = - \lim_{\tau\to\infty} \frac{1}{\tau} \log p(\bbar|\tau)  \; .
\label{equ:rate-fn}
\eeq
As anticipated in Sec.~\ref{sec:intro-fluct}, the interesting case is where this limit is finite (and non-zero), so the relevant fluctuations occur with probabilities that decay exponentially with $\tau$.
In this case we say that $\bbar_\tau$ obeys a large deviation principle (LDP) and $I$ is called the \emph{rate function}.\footnote{%
The mathematical theory of large deviations~\cite{denH-book} expresses LDPs in a more general way that involves probabilities of events instead of probability densities, and also places some additional restrictions on rate functions.  The details of the mathematical theory 
can be important in some physical situations, but we concentrate here on simple cases for which the presentation given here is adequate.
}
The rate function is non-negative, $I(b)\geq 0$ for all $b$.
In cases where an LDP holds with rate function $I$, we write
\beq
p(\bbar|\tau) \asymp \ee^{-\tau I(\bbar) } \; .
\label{equ:asymp}
\eeq
The meaning of the asymptotic equality symbol $\asymp$ is that (\ref{equ:asymp}) is equivalent to (\ref{equ:rate-fn}), see~\cite{Ellis1995}.
It is a general property of LDPs that the argument of the exponential in (\ref{equ:asymp}) is the product of the rate function and a large parameter that is called the \emph{speed} of the LDP.
In (\ref{equ:rate-fn},\ref{equ:asymp}) the speed is $\tau$, which is an assumption of the theory presented so far.  There are physical systems where time-averaged quantities obey LDPs with other speeds (for example~\cite{Harris2009,Krapivsky2014,Harris2015,Nickelsen2018}) but we focus here on  LDPs with speed $\tau$, which is the most common situation.

\begin{figure}
\includegraphics[width=80mm]{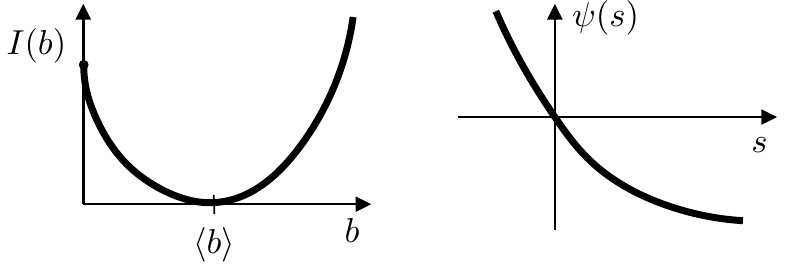}
\caption{Sketch of a rate function and an SCGF for a positive quantity $\bbar$. In this example, both functions are convex and related by Legendre transformation; the rate function has a single minimum at the ensemble averarge $\langle b\rangle$; the SCGF has $\psi(0)=0$ and $\psi'(0)=-\langle b\rangle$, it is monotonic because $\bbar$ is positive.}
\label{fig:simple-rate}
\end{figure}

In simple cases (see Sec~\ref{sec:scope} for examples), 
the rate function $I$ is analytic and strictly convex, with a 
unique minimum at $b=\langle b\rangle$, and $I(\langle b \rangle)=0$.
In this case~\cite{denH-book,Touchette2009}, $\bbar_\tau$ obeys a 
a central limit theorem (as in~\cite{Dechant2011}), with a variance $\sigma^2_b/\tau$ that is related to the curvature of the rate function  as
$\sigma_b^2 = 1/I''(\langle b\rangle)$.
The next step is to define the \emph{scaled cumulant  generating function} (SCGF),
\beq
\psi(s) = \lim_{\tau\to\infty}\frac{1}{\tau}  G(s,\tau) \; .
\label{equ:psi}
\eeq
One sees  from (\ref{equ:cgf}) that
$\psi'(0)=-\langle b\rangle$ and $\psi''(0)=\sigma^2_b$.  

The rate function and the SCGF are related by a Legendre transformation:
\beq
\psi(s) = \sup_b [ -s b - I(b) ] \; .
\label{equ:psi-legendre}
\eeq
This is (a particular case of) Varadhan's lemma~\cite{denH-book}. It can be motivated by writing (\ref{equ:cgf}) as 
\beq
G(s,\tau) = \log \int p(\bbar|\tau) \ee^{-s\tau\bbar} \mathrm{d}\bbar \;,
\eeq
and substituting (\ref{equ:asymp}), then doing the integral by the saddle-point method.
If the function $\psi$ is analytic then one has also
\beq
I(b) = \sup_s [ -s b - \psi(s) ] \; .
\label{equ:I-legendre}
\eeq
An important result in large deviation theory is the G\"artner-Ellis theorem~\cite{denH-book,Touchette2009}: it allows large deviation results like (\ref{equ:asymp}) to be proved, as long as the SCGF $\psi$ obeys certain conditions.  In such cases the rate function can then be derived from (\ref{equ:psi-legendre}).

A schematic illustration of two  functions $I$ and $\psi$ related by Legendre transform are shown in Fig.~\ref{fig:simple-rate}.
Note, the minus sign in (\ref{equ:cgf}) means that the behaviour of the SCGF for $s>0$ is relevant for the rate function for $b<\langle b\rangle$, and vice versa.

\subsection{Applicability of the theory}
\label{sec:scope}

The  theory of Sec.~\ref{sec:def} can be applied to a wide range of models, but some assumptions are required in order to ensure that the limit in (\ref{equ:rate-fn}) is finite and non-zero. Some results including (\ref{equ:I-legendre}) also rely on analytic properties of $\psi$.
In this article, much of our analysis is based on two main classes of system, which are Markov chains and diffusion processes.
We make several assumptions, which ensure that the models are ergodic, the limit (\ref{equ:rate-fn}) is well-behaved, and the functions $I$ and $\psi$ are analytic and strictly convex, as discussed in~\cite{Chetrite2015}.
These cases are useful to illustrate the theory.  However, the tools of large-deviation theory are not at all restricted to these cases; this will become clear in later sections.

\subsubsection{Finite-state Markov chains.}  We consider finite Markov chains, so the configurations $\CC$ come from a finite set.   
They may evolve in either continuous- or discrete-time.  In continuous time, a model is defined by specifying the transition rates between the configurations, which are denoted by $W(\CC\to\CC')$.   Prominent examples in this case include exclusion processes and Ising-like models on finite lattices.  In this case $\bbar$ may be defined as a time integral as in (\ref{equ:bbar}), or one may take the more general form~\cite{Garrahan2009}
\beq
\bbar_\tau = \frac{1}{\tau}\sum_{{\rm jumps}\; \CC\to\CC'} \alpha(\CC,\CC') + \frac{1}{\tau}\int_0^\tau h(\CC_t) \mathrm{d}t 
\label{equ:bbar-alpha}
\eeq
where the functions $\alpha,h$ correspond to observable quantities similar to $b$ in (\ref{equ:bbar}), and the sum runs over the transitions that take place in the trajectory.  
This type of time-averaged quantity  is particularly useful when considering time-averaged currents: for example, if the model involves particles hopping on a $1d$ lattice with periodic boundaries, one may take $\alpha=1$ for jumps to the right and $\alpha=-1$ for jumps to the left~\cite{Derrida2007}, with $h=0$.

If the continuous-time Markov chain is finite and irreducible and $\alpha,h$ are finite then the limits in (\ref{equ:rate-fn},\ref{equ:psi}) certainly exist, and the functions $I$ and $\psi$ are analytic and strictly convex~\cite{Chetrite2015}. 
In discrete time the situation is the same, as long as the Markov chain is also aperiodic.

\subsubsection{Diffusion processes.}  

We also consider models defined by stochastic differential equations (or Langevin equations).   In this case the configurations $\CC$ are vectors in $d$-dimensional space and they evolve by 
\beq
\mathrm{d}\CC_t = v(\CC_t) \mathrm{d}t + \sqrt{2}\sigma(\CC_t) \circ \mathrm{d}W_t
\eeq
where the circle indicates a Stratonovich product.  Here, $v$ is a vector-valued drift, $\sigma$ is a matrix-valued noise strength, and $W_t$ is a $d$-dimensional standard Brownian motion. For models in this class, some technical restrictions are needed on the functions $v,\sigma$, in order to establish existence of the limits in (\ref{equ:rate-fn},\ref{equ:psi}) and convexity of the rate function.  For simplicity, we restrict to systems defined on finite domains, with periodic boundary conditions.  In this case it is sufficient that $v,\sigma$ should be finite and the matrix $\sigma\sigma^\dag$ should not have any zero eigenvalues.
In this case $\bbar$ may again be defined as in (\ref{equ:bbar}), or one may consider~\cite{Chetrite2015}
\beq
\bbar_\tau = \frac{1}{\tau} \int_0^\tau a(\CC_t) \circ \mathrm{d}\CC_t +  \frac{1}{\tau}\int_0^\tau h(\CC_t) \mathrm{d}t
\label{equ:bbar-aa}
\eeq
where now $a,h$ are vector-valued and scalar functions respectively.  
A simple case takes $a$ to be constant and $h=0$ in which case $\bbar_\tau$ is a time-averaged current in the direction $a$.  (In systems with closed boundaries then such time-averaged currents must vanish as $\tau\to\infty$ but periodic systems can support trajectories with sustained non-zero current.)
The functions $b,a,h$ are all assumed to be finite.  


\subsection{Analogy between $\tau\to\infty$ and thermodynamic limit}
\label{sec:thermo}

\newcommand{\kB}{k_{\rm B}}

This article focusses on large deviations of time-averaged quantities, but there are other situations where large deviation theory is relevant in physics.  The most prominent example is the theory of the thermodynamic limit~\cite{Ellis-book,Ruelle2004}.  We briefly outline the analysis of this limit within large deviation theory, which motivates an analogy between large-time limits and thermodynamic limits.  
A more detailed discussion of this analogy is given in the review of Touchette~\cite{Touchette2009}, see also~\cite{Gaspard1995,Lecomte2007,Garrahan2009}.
The analogy is useful for two reasons.  First, it provides valuable intuition about dynamical large deviations, since thermodynamic theories may be more familiar than dynamical ones.  Second, it provides a route whereby established methods from thermodynamics can be generalised, in order to address dynamical problems.

Within the analogy, the CGF in (\ref{equ:cgf}) corresponds to a \emph{difference in free-energy} between two states.  Specifically, consider a thermodynamic system of volume $V$, where the energy  of configuration $\CC$ is ${\cal E}(\CC)$.
Define $\beta=(\kB T)^{-1}$ where $\kB$ is Boltzmann's constant, so the Boltzmann distribution of this system is $p_\beta(\CC)=\ee^{-\beta {\cal E}(\CC)}/Z_\beta$, where $Z_\beta$ is the partition function.  We denote averages with respect to $p_\beta$ by $\langle \cdot\rangle_{\rm Boltz}$.

Now consider a perturbation to this system where the (extensive) energy is modified by $\Delta{\cal E}(\CC)=-hVm_V(\CC)$.  For example, $m_V$ might be the (intensive) magnetisation of an Ising model, and $h$ its conjugate (magnetic) field.
The free energy difference between the original system and this new state is $\Delta F$.  It satisfies
\beq
-\beta \Delta F(\beta,h,V) = \log \big\langle \ee^{\beta h V m_V(\CC) } \big\rangle_{\rm Boltz}
\eeq
which is analogous to (\ref{equ:cgf}) with $(-\beta h,V,m_V)\to(s,\tau,\overline{b}_\tau)$ and $(-\beta\Delta F) \to G$.
For the purposes of this analogy we consider $\beta$ to be a fixed number, 
it is the field $h$ that is the analogue of the parameter $s$ from Sec.~\ref{sec:def}.  In that Section we considered the limit $\tau\to\infty$, here we consider $V\to\infty$.

Application of large-deviation theory to the fluctuations of $m_V$ requires that this quantity is intensive, which means that it can be expressed as an average over the (large) system, analogous to the time-average in (\ref{equ:bbar}).  Then standard thermodynamic arguments for large systems imply that $\Delta F$ is extensive: $\Delta F(\beta,h,V)\approx V \Delta f(\beta,h)$, where $\Delta f(\beta,h)$ is a difference in \emph{free-energy density}.  Comparing with  (\ref{equ:psi}), the dynamical SCGF $\psi(s)$ is analogous to $-\beta \Delta f(\beta,h)$.  Continuing the analogy shows that $m_V$ for the unperturbed system has a probability distribution
\beq
p(m|V,\beta) \asymp \ee^{-V I(m,\beta) }
\eeq
with $I(m,\beta) = \sup_h [ \beta hm + \beta\Delta f(\beta,h) ]$, similar to (\ref{equ:asymp},\ref{equ:I-legendre}).  
Just like the dynamical case, some care is required with this analysis in cases where $\Delta f(\beta,h)$ is not analytic.  
These cases correspond to thermodynamic phase transitions, for which there is a well-developed theory: see for example~\cite{Ellis-book,sokal93}.  Sec.~\ref{sec:phase-transition} discusses some ways that the thermodynamic theory of phase transitions can be generalised to the dynamical context.

\subsection{Biased ensembles of trajectories (the $s$-ensemble)}
\label{sec:sens}

In the thermodynamic setting, it is natural to consider a family of Boltzmann distributions, parameterised by $h$.  
We now introduce corresponding distributions for trajectories, which
we refer to as $s$-ensembles~\cite{Garrahan2009} or biased ensembles~\cite{Chetrite2015}.
Let $\bm{\CC}$ indicate a trajectory of the system of interest, where the time $t$ runs from $0$ to $\tau$.  
This trajectory has a probability density $P_\tau(\bm{\CC})$, which has the property that $\langle F \rangle = \int F(\bm{\CC}) P_\tau(\bm{\CC})  \mathrm{d}\bm{\CC}$.\footnote{%
It is not trivial to define the integration measure $\mathrm{d}\bm{\CC}$, but see~\cite{Garrahan2009} for an explicit construction for finite Markov chains.
A more rigorous mathematical approach would sidestep this problem by working directly with probability measures for trajectories.
The analysis of this work can be reformulated in that way: one should replace integration measures
$P_\tau(\bm{\CC})  \mathrm{d}\bm{\CC}$ by $\mathrm{d} P_\tau(\bm{\CC})$ and ratios of probability densities $P(\bm{\CC})/Q(\bm{\CC})$ by
Radon-Nikodym derivatives ${\rm d}P/{\rm d}Q$.   All conclusions remain unchanged.
}
Note that the probability of the initial state $\CC_0$ is included in $P_\tau(\bm{\CC)}$.    

The probability
density for trajectory $\bm{\CC}$ in the biased ensemble is
\beq
P^s_\tau( \bm{\CC} ) =  P_\tau( \bm{\CC} ) \ee^{-s\tau \bbar_\tau(\bm{\CC}) -  G(s,\tau)  }
\label{equ:Ps}
\eeq
which is normalised, by (\ref{equ:cgf}).  
The average of
any trajectory-dependent observable $F$ within this ensemble is
\beq
\langle F \rangle_s = \frac{ \langle F \ee^{-s\tau\overline{b}_\tau} \rangle }{ \langle \ee^{-s\tau\overline{b}_\tau} \rangle } \; .
\label{equ:s-ens}
\eeq
Note that these averages depend implicitly on 
the trajectory length $\tau$.

In the analogy with thermodynamics, (\ref{equ:Ps}) corresponds to a Boltzmann distribution, in the canonical ensemble.  
As discussed in~\cite{Garrahan2009}, standard thermodynamic 
arguments for equivalence of ensembles then indicate that typical trajectories of (\ref{equ:Ps}) should be similar
to typical trajectories from an associated microcanonical ensemble, where the value of $\bbar_\tau$ is constrained to a specific value.
A precise characterisation of this ensemble-equivalence is given in Refs.~\cite{Chetrite2013,Chetrite2015}.  

An important observation is that the initial and final conditions of the trajectory are analogous to boundaries of thermodynamic systems, where the behaviour may differ from the bulk.  Thermodynamic equivalence of ensembles applies to observable quantities that are evaluated in finite regions, within the bulk of a large system.  In biased ensembles, these correspond to observables that are well-separated (in time) from the initial and final conditions at $t=0,\tau$.  

Bearing this mind, it is useful to consider a one-time dynamical observable $a(\CC_t)$, such as the instantaneous energy of the system $E(\CC_t)$.  This quantity is associated with two different probability distributions, depending on the time $t$~\cite{Bodineau2008,Garrahan2009,Jack2010}.  The bulk is characterised by a distribution which we define by evaluating the observable at a randomly-chosen time:
\beq
P_{\rm ave}(a|s) = \lim_{\tau\to\infty} \frac{1}{\tau} \int_0^\tau \big\langle \delta[ a - a(\CC_t) ] \big\rangle_s \mathrm{d}t \; .
\eeq
Alternatively one may evaluate the same observable at the final time $\tau$ to obain
\beq
P_{\rm end}(a|s) = \lim_{\tau\to\infty} \langle \delta[ a - a(\CC_\tau) ] \rangle_s \; .
\label{equ:P-end}
\eeq
The presence of boundaries means that $P_{\rm ave}\neq P_{\rm end}$ in general.  
In the cases that we consider, the bulk of the $s$-ensemble is time-translation invariant (similar to homogeneity of thermodynamic systems),
which means that $P_{\rm ave}$ can also be evaluated as $P_{\rm ave}(a|s) =  \lim_{\tau\to\infty} \langle \delta[ a - a(\CC_{u\tau}) ] \rangle_s$
for any $u$ with $0<u<1$.\footnote{%
The inequalities are strict so $u=0,1$ are excluded, in particular taking $u=1$ recovers $P_{\rm end}$.
}

\subsection{Formulation as eigenproblem (operator approach)}
\label{sec:spectral} 

We describe two types of theoretical approach by which results for large deviations can be obtained.  
This section describes the first method, which is to characterise $\psi(s)$ as the largest eigenvalue of an operator (or matrix), which is called a tilted generator or a biased master operator.   In the physical context, this was the approach applied (for stochastic models) in~\cite{Derrida1998,Lebowitz1999}, see also~\cite{Gwa1992,Eyink1996,Gaspard1998}.
To explain it, 
define
\beq
\rho(\CC | s,\tau) = \left\langle \ee^{-s \tau \bbar_\tau} \delta( \CC - \CC_\tau ) \right\rangle
\label{equ:rho-s-C}
\eeq
where the delta function restricts the average to trajectories that end in state $\CC$.   Comparing with  (\ref{equ:cgf}), one sees that $G(s,t) = \log \int \rho(\CC|s,\tau) \mathrm{d}\CC$.

The time derivative of $\rho$ behaves as
\beq
\frac{ \partial }{ \partial \tau  } \rho(\CC | s,\tau) = {\cal W}_s \rho(\CC | s,\tau)
\label{equ:L-tilt}
\eeq
where ${\cal W}_s$ is an $s$-dependent linear operator.\footnote{%
This is a tilted version of what would be called in mathematics the \emph{forward generator}, the `tilting' refers to the effect of $s$ and setting $s=0$ recovers the usual (forward) generator.  The adjoint of ${\cal W}_s$ is the (tilted) backward generator.  Mathematical analyses are typically framed in terms of the backwards generator.} 
For example, in finite-state Markov chains (with $n$ states) then 
\beq
{\cal W}_s \rho(\CC | s,\tau) =  \sum_{\CC'} M_s(\CC,\CC') \rho(\CC'|s,\tau) 
\label{equ:LM}
\eeq
where $M_s$ is a matrix of size $n\times n$ that depends on the transition rates of the model and on the observable $\overline{b}_\tau$~\cite{Derrida2007,Garrahan2009,Chetrite2015}.
For diffusion processes then ${\cal W}_s$ is an operator that involves first and second derivatives with respect to $\CC$,
an example is given in (\ref{equ:WW-simple}), below.

The large-time behavior of the solution of (\ref{equ:L-tilt}) can be deduced by considering the largest eigenvalue of ${\cal W}_s$.  [In the example of (\ref{equ:LM}), this is simply the largest eigenvalue of $M$.]  Anticipating the answer, we assume that this largest eigenvalue is unique and we denote it by $\psi(s)$.  The associated eigenvector (or eigenfunction) is $P_{\rm end}(\CC|s)$ which we define to be normalised as a probability distribution $\int P_{\rm end}(\CC|s)\mathrm{d}\CC=1$.  So the eigenproblem is
\beq
{\cal W}_s  P_{\rm end} = \psi(s) P_{\rm end}
\label{equ:eigen-Ls}
\eeq
and the solution of (\ref{equ:rho-s-C}) is
\beq
\rho(\CC | s,\tau) =  \ee^{\psi(s)\tau} P_{\rm end}(\CC|s) [ A_s + O(\ee^{-\tau\Delta}) ]
\label{equ:rho-spectral}
\eeq
for some constant $A_s$ (independent of $\tau$).  In the correction term, $\Delta$ is the gap between the largest and second-largest eigenvalues of ${\cal W}_s$.\footnote{%
For the cases described in Sec.~\ref{sec:scope}, the gap $\Delta$ is strictly positive.  Models (and limits) where $\Delta$ vanishes are often associated with anomalous fluctuations, including dynamical phase transitions, see Sec.~\ref{sec:phase-transition}.}
Integrating over $\CC$ one sees that $G(s,\tau) \approx \tau\psi(s) + \log A_s$, consistent with (\ref{equ:psi}).
By (\ref{equ:s-ens}), we also identify $\rho(\CC | s,\tau) \ee^{-G(s,\tau)}$ with $\langle \delta( \CC - \CC_\tau ) \rangle_s$.
Taking $\tau\to\infty$ one sees from (\ref{equ:rho-spectral},\ref{equ:psi}) that it is consistent to identify the eigenvector of ${\cal W}_s$ with $P_{\rm end}$
as defined in (\ref{equ:P-end}).

Note that the operator ${\cal W}_s$ is not generally Hermitian (self-adjoint).  The eigenvector that we identified here as $P_{\rm end}$ is the right eigenvector.  The role of the left eigenvector will be discussed in the next section.

To summarise, 
large deviations of $b_\tau$ can be characterised by analysing the properties of the tilted operator ${\cal W}_s$, as in~\cite{Derrida1998,Lebowitz1999}.  
This approach is valuable as a tool for explicit computations (especially in finite-state Markov chains where the matrix $M$ is finite).
In addition, it establishes a connection between large-deviation problems and eigenproblems that are familiar from quantum mechanics.  
Like the analogy with thermodynamics discussed above, this connection with quantum mechanics is useful in practice because it means that methods from that field can be generalised in order to analyse large deviations~\cite{Jack2010,Gwa1992,Elmatad2010,banuls2019}.

\subsection{Control representation and auxiliary process}
\label{sec:control}

This section describes a second method for analysis of large deviations, based on optimal control theory~\cite{bertsekas-book}.  
One advantage of this method is that it is built on a variational formula, which can be very useful for deriving approximate results in situations where diagonalisation of ${\cal W}_s$ is not possible.
The method has a transparent physical interpretation which is that (rare) large deviation events can be characterised by deriving a new physical model whose \emph{typical trajectories} resemble closely the rare events of interest.  This new model is called here the \emph{optimally controlled process}, following earlier work by Fleming~\cite{Fleming85} and (more generally) the book of Dupuis and Ellis~\cite{dupuis-book}.  In previous work it has been called a driven process~\cite{Chetrite2015,Chetrite2015var} or an auxiliary process~\cite{Jack2010,Jack2015b}, see also~\cite{Gwa1992,Maes2008,Simha2008,Simon2009}.

 Consider first a general controlled process (not necessarily optimal).  Let $\langle F\rangle_{\rm con}$ denote the average of a path-dependent quantity $F$, in this process.  The probability density for trajectories in the controlled process is $P^{\rm con}_\tau(\bm{\CC})$.
Then a useful general formula~\cite[Prop. 1.4.2]{dupuis-book} is 
\beq
G(s,\tau) \geq -s\tau \langle \bbar_\tau \rangle_{\rm con} - {\cal D}(P^{\rm con}_\tau || P_\tau)
\label{equ:G-var}
\eeq
where 
\beq
 {\cal D}(P^{\rm con}_\tau || P_\tau) = \left\langle \log \frac{ P^{\rm con}_\tau(\bm{\CC}) }{ P_\tau(\bm{\CC}) }  \right\rangle_{\rm con} 
 \label{equ:KL}
\eeq
is the Kullback-Leibler (KL) divergence between $P^{\rm con}$ and $P$.\footnote{%
In a more rigorous approach, the ratio of probability densities in this definition would be replaced by a Radon-Nikodym derivative.
}
The KL divergence is non-negative and is equal to zero only if $P^{\rm con}_\tau=P_\tau$.  In the thermodynamic setting of Sec.~\ref{sec:thermo}, Equ.~(\ref{equ:G-var}) is the Gibbs-Bogoliubov inequality, see~\cite{Falk1970}, in particular their Equ.~(25).
It is possible to find a controlled process where (\ref{equ:G-var}) becomes an equality.
To see this, use (\ref{equ:s-ens},\ref{equ:KL}) to rewrite the right hand side of (\ref{equ:G-var}):
\beq
-s\tau \langle \bbar_\tau \rangle_{\rm con} - {\cal D}(P^{\rm con}_\tau || P_\tau) = G(s,\tau) -  {\cal D}(P^{\rm con}_\tau || P^s_\tau) \; .
\eeq
Hence equality is possible in (\ref{equ:G-var}) only if $P^{\rm con}_\tau = P^s_\tau$: the controlled process must reproduce the probability distribution of the $s$-ensemble.\footnote{%
It is not trivial to construct a stochastic process whose probability distribution of trajectories achieves  $P^{\rm con}_\tau=P^s_\tau$, this is related to the theory of dynamic programming~\cite{bertsekas-book}. The construction of such a process is possible for all examples considered here, although the controlled process may be complicated. For example, its transition rates may depend on time, see for example~\cite{Chetrite2015}.}

The bound (\ref{equ:G-var}) can be analysed using tools from stochastic optimal control theory~\cite{bertsekas-book}, see also~\cite{Chetrite2015var}.
The general aim of this theory is to find (controlled) Markov processes that maximise (or minimise) quantities like the right hand side of (\ref{equ:G-var}), which are interpreted as cost functions.
For example, the process $P^{\rm con}$ might consist of requests which arrive randomly in a queue, and a stochastic rule for dealing with these requests.  In this case a suitable cost would be some combination of the mean waiting time in the queue and the resource required to implement the policy.  One seeks the policy that minimises the cost.  Such problems have been studied in detail, they are obviously applicable in practical settings and they are also mathematically tractable~\cite{bertsekas-book}.

Returning to the large-deviation context, observe that computation of the large-deviation rate function does not require a full characterisation of $G(s,\tau)$ but only of $\psi(s)$, which is related to $G(s,\tau)$ by (\ref{equ:psi}).  Hence 
\beq
\psi(s) \geq \lim_{\tau\to\infty} \left[ -s \langle \bbar_\tau \rangle_{\rm con} - \frac{1}{\tau} {\cal D}(P^{\rm con}_\tau || P_\tau) \right] \; .
\label{equ:psi-var}
\eeq
A key observation is that for the standard cases of Sec.~\ref{sec:scope},  equality can 
be achieved in this formula by an (optimally)-controlled process that is Markovian and stationary~\cite{Jack2010,Chetrite2015,Nemoto2011}.  This is a very useful simplification.
From a comparison with (\ref{equ:psi-legendre}), one may expect that 
$
I(b) =  \tau^{-1} {\cal D}(P^{\rm con}_\tau || P_\tau),
$ 
where $P^{\rm con}$ is a controlled process with $\langle \bbar_\tau\rangle_{\rm con}=b$.  In this case,
\beq
I(b ) = \inf \left[  \lim_{\tau\to\infty} \frac{1}{\tau} {\cal D}(P^{\rm con}_\tau || P_\tau) \right]
\label{equ:Ib-KL}
\eeq
where the minimisation is over stationary Markovian controlled processes for which $\langle \bbar_\tau\rangle_{\rm con}=b$.\footnote{%
The results (\ref{equ:G-var},\ref{equ:psi-var}) are extremely general but (\ref{equ:Ib-KL}) is similar to (\ref{equ:I-legendre}) in that it requires assumptions related to analyticity and convexity of $\psi$ and $I$.  These assumptions are valid for models within the scope of Sec.~\ref{sec:scope}.%
}

The final result (\ref{equ:Ib-KL}) has an intuitive interpretation, it states that the \emph{least unlikely} mechanism for achieving a rare event with $\bbar_\tau  = b$ can be reproduced by a controlled process that minimises the KL divergence.  A central idea of large-deviation theory~\cite{denH-book} is that this least unlikely mechanism is sufficient to characterise the rare event.  The variational principle means that the controlled process differs as little as possible from original process; the size of the difference is quantified via the KL divergence.

\subsection{Equivalence of different large deviation problems}
\label{sec:gauge}

An interesting aspect of the theory presented here is that the same optimally-controlled process may appear as 
the solution to several different large deviation problems.  In the operator formalism, 
this happens because the same operator ${\cal W}_s$ may appear in several different contexts. 
In fact, this is a very common situation.
To see the reason, we define 
\beq
  \overline{g}_\tau(\bm{\CC}) = \frac{1}{\tau} \log \frac{ P^{\rm con}_\tau(\bm{\CC}) }{ P_\tau(\bm{\CC}) }  \; .
 \label{equ:rk-g}
\eeq
For models in the scope of Sec.~\ref{sec:scope}, the quantity $\overline{g}_\tau$ has a representation as either (\ref{equ:bbar-alpha}) or (\ref{equ:bbar-aa}). 
Hence one sees that the biased ensemble $P^s_\tau$ of (\ref{equ:Ps}) can also be characterised as a biased ensemble for the controlled process:
\beq
P^s_\tau( \bm{\CC} ) \propto  P_\tau^{\rm con}( \bm{\CC} ) \ee^{-s\tau \bbar_\tau(\bm{\CC}) - \tau \overline{g}_\tau(\bm{\CC})  }
\label{equ:Ps-con}
\eeq
Given a biased ensemble of interest, one may choose the controlled process (and hence $\overline{g}$) in order to 
transform the problem into a form that is more tractable.  This is very useful for numerical work~\cite{Nemoto2016,Nemoto2017first,Ray2018,Brewer2018,Nemoto2019}.  It also enables analytic progress.  For example, in biased ensembles where $\overline{b}_\tau$ is of the form given in (\ref{equ:bbar-alpha}) of (\ref{equ:bbar-aa}), it is simple to construct a controlled process such that the quantity 
$s\bbar_\tau(\bm{\CC}) + \overline{g}_\tau(\bm{\CC})$ that appears in (\ref{equ:Ps-con}) reduces to a simple time-integral as in (\ref{equ:bbar}).
Hence biased ensembles $P_s$ with $\overline{b}_\tau$ as in (\ref{equ:bbar-alpha}) have alternative formulations where the dynamics is modified but the bias has the (simpler) form (\ref{equ:bbar}).  
This observation was used in~\cite{Garrahan2009} to relate large deviations of the dynamical activity in spin models to large deviations of the time-integrated escape rate, see also~\cite{Jack2015} which discusses some relationships between large deviations of currents and dynamical activities.

\subsection{Connection of operator and optimal-control approaches}

There is a deep connection between the optimal control approach of Sec.~\ref{sec:control} and the operator approach of Sec.~\ref{sec:spectral}.
A similar connection appears in quantum mechanics, where one may use either an operator approach
or an approach based on  path integrals.  

A general method to connect operator equations and controlled processes is to 
maximise the right hand side of (\ref{equ:psi-var}) over some class of controlled processes,
in order to find an optimally-controlled model.
This variational problem is equivalent to solving for the largest eigenvalue of an operator
${\cal W}_s^\dag$, which is the Hermitian conjugate (adjoint) of the operator ${\cal W}_s$ discussed above.  (The eigenvalue appears as the value of a Lagrange multiplier.)
We present an example calculation for a simple diffusion process, after which we summarise the resulting general picture.

Consider large deviations of $\bbar_\tau$ as in (\ref{equ:bbar}), for a diffusion problem described by a stochastic differential equation with additive noise:
\beq
\mathrm{d}\CC_t = v(\CC_t) \mathrm{d}t + \sqrt{2} \, \mathrm{d}W_t \; ,
\label{equ:sde-simple}
\eeq
where $\CC_t$ is a $d$-dimensional vector and $W_t$ a $d$-dimensional standard Brownian motion.  
Using the operator method, the SCGF can be obtained for this process by solving the eigenvalue problem
\begin{align} 
\psi(s) P_{\rm end} & = {\cal W}_s P_{\rm end} 
\nonumber\\
& =  \nabla^2 P_{\rm end} - \nabla\cdot (v P_{\rm end} ) -  s b P_{\rm end}  \; .
\label{equ:WW-simple}
\end{align}
(The second line is an explicit formula for ${\cal W}_s$, the derivatives are with respect to $\CC$.)

The controlled process is obtained from (\ref{equ:sde-simple}) by replacing $v$ with $(v-\nabla\phi)$ where $\phi$ is a control-potential, that is
\beq
\mathrm{d}\CC_t = [v(\CC_t)-\nabla\phi(\CC_t)] \mathrm{d}t + \sqrt{2} \, \mathrm{d}W_t \; .
\label{equ:sde-con}
\eeq
Similarly to~\cite{Nemoto2011,Chetrite2015var}, we show in Appendix~\ref{app:control} that if this control potential is used with (\ref{equ:psi-var}), maximising the resulting bound on $\psi$ is equivalent to solving the eigenproblem (\ref{equ:WW-simple}).  In particular, the optimal control 
may be expressed as $\phi = -2\log {\cal F}$ where ${\cal F}$ solves the eigenproblem
\begin{align} 
\psi {\cal F} 
& = {\cal W}_s^\dag {\cal F} \; ,
\label{equ:WWdag-simple}
\end{align}
in which ${\cal W}_s^\dag$ is the Hermitian conjugate (adjoint) of the operator ${\cal W}_s$ given in (\ref{equ:WW-simple}).  Its form is given in (\ref{equ:F-W}).
Equ.~(\ref{equ:WWdag-simple}) is an eigenproblem for ${\cal W}_s^\dag$, whose largest eigenvalue was already shown to be the SCGF $\psi$.\footnote{
Of course, ${\cal W}_s^\dag$ and ${\cal W}_s$ have the same eigenvalues.}
Constructing the controlled process from the corresponding eigenfunction ${\cal F}$ achieves equality in (\ref{equ:psi-var}) -- hence this is an optimally-controlled process.

The conclusion of this analysis is that solving the eigenvalue problem (\ref{equ:psi-var-sde}) is equivalent to optimising (\ref{equ:psi-var}) over controlled processes of the form (\ref{equ:sde-con}).  Also, the optimal control potential and the eigenvector are related as ${\cal F}=\ee^{-\phi/2}$.  So the same information is available by the operator and optimal-control approaches.

We have analysed the simple model (\ref{equ:sde-simple}) but this structure is very general, see also~\cite{dupuis-book}.  
Analogous steps can be applied to all the models of Sec.~\ref{sec:scope}.  Taking $\overline{b}_\tau$ as in (\ref{equ:bbar}) it is sufficient in these cases to consider controlled processes that are obtained by adding conservative control forces, as the derivative of a potential.  
For Markov chains with transition rates $W(\CC\to\CC')$, the appropriate controlled dynamics is~\cite{Maes2008,Jack2010}
\beq
W^{\rm con}(\CC\to\CC') = \ee^{\phi(\CC)/2} W(\CC\to\CC') \ee^{-\phi(\CC')/2} \; .
\eeq
For $\overline{b}_\tau$ as in (\ref{equ:bbar-alpha},\ref{equ:bbar-aa}) one should first use the method of Sec.~\ref{sec:gauge} to transform the problem to a form where $\overline{b}_\tau$ has the form given in (\ref{equ:bbar}): this may require a non-conservative control force.  One then adds an additional conservative control force, as the gradient of $\phi$.  It is sufficient to optimise over this $\phi$.

We end this Section by observing that for time-reversal symmetric systems, both the eigenvalue problem and the optimal-control problem can be simplified.  If 
(\ref{equ:sde-simple}) represents an equilibrium (time-reversal symmetric) system then $v=-\nabla U$ for some potential $U$, so the controlled system (\ref{equ:sde-con}) is also time-reversal symmetric (with potential $U+\phi$).  In this case the steady state of the controlled system is a Boltzmann distribution $\mu \propto \ee^{-(U+\phi)}$.   Then (\ref{equ:G-var}) yields a simple variational result
\beq
\psi(s) = \sup_{\cal F} \frac{ \int \ee^{-U} ({\cal F} {\cal W}_s^\dag {\cal F})  \mathrm{d}\CC }{ \int \ee^{-U} {\cal F}^2 \mathrm{d}\CC }  \; ,
\label{equ:psi-var-hermitian}
\eeq
which is equivalent to the Rayleigh-Ritz formula for the largest eigenvalue of a self-adjoint operator,  see also~\cite{Eyink1996}.  The physical origin of this simplification is the time-reversal symmetry of the biased ensemble (\ref{equ:s-ens}).
The ${\cal F}$ that maximises the right hand side of (\ref{equ:psi-var-hermitian}) is the eigenfunction of ${\cal W}_s^\dag$ and gives the optimal control potential as $\phi = -2\log {\cal F}$.

\section{Dynamical phase transitions}
\label{sec:phase-transition}

We emphasised in Sec.~\ref{sec:scope} that finite systems are typically associated with analytic rate functions and SCGFs.
However, there are many examples of rate functions that have singularities.
For example, this can occur in Markov chains with infinite state spaces~\cite{Harris2005,Bodineau2005,Garrahan2007,Jack2010}, which are not covered by Sec.~\ref{sec:scope}.
Motivated by the analogy with thermodynamics discussed in Sec.~\ref{sec:thermo}, these singularities can be identified as phase transitions.

Physically, the key feature is that singularities are (usually) associated with  a qualitative difference in 
mechanism between rare events with different values of $\bbar_\tau$.
There are several situations in which such behaviour can arise.  We focus here on one broad class of phase transitions,
which we describe as \emph{space-time phase transitions}, sometimes called \emph{trajectory phase transitions}~\cite{Garrahan2007}, see also~\cite{Mackay2008}.  These occur in large systems where the observable $b$ in (\ref{equ:bbar}) is an intensive variable in the spatial (thermodynamic) sense, see below.   

Other kinds of dynamical phase transition have also been discussed in the context of  dynamical large deviations~\cite{Harris2005,Bodineau2005,Bunin2012,Suri2014,Nyawo2016,Jack2019-growth}.  
Those results show that singular rate functions can occur for a variety of different reasons.  They also show that systems outside the scope of Sec.~\ref{sec:scope} cannot be assumed to have analytic rate functions, even if the models appear very simple.

\subsection{Thermodynamics in space-time}
\label{sec:spacetime-general}

\begin{figure}
    \includegraphics[width=80mm]{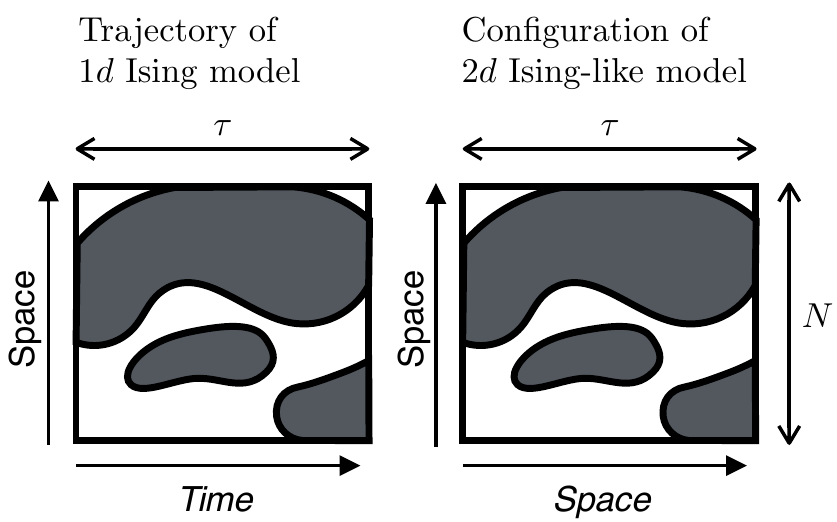}
    \caption{The idea of space-time thermodynamics~\cite{Merolle2005,Jack2006-spacetime,Garrahan2009} is that dynamical trajectories of $d$-dimensional models (left) can be analysed by mapping them to configurations of ($d+1$) thermodynamic models (right).  The thermodynamic system has size $N\times \tau$ and one analyses its behaviour in the joint limit $N,\tau\to\infty$.  }
    \label{fig:spacetime}
\end{figure}

In Sec.~\ref{sec:thermo} we described an analogy between large deviations of time-averaged quantities and the thermodynamic limit.
In this Section we are concerned with large deviations of time-averaged quantities in large systems.  
As a guiding example, we consider the one-dimensional Ising model with  periodic boundaries, evolving by Glauber dynamics, as in~\cite{Jack2010}.  There are $N$ spins and the state of the $i$th spin at time $t$ is $\sigma_{i,t}=\pm1$. 
We consider a joint limit 
of large time $\tau\to\infty$ and large system size $N\to\infty$.  

To analyse this situation it is useful to make a mapping between trajectories of a $d$-dimensional model and configurations of a corresponding $(d+1)$-dimensional thermodynamic system~\cite{Gwa1992,Lecomte2007,Garrahan2009,Merolle2005,Jack2006-spacetime,Maes1999}.   
The key idea is that the time $t$ in the dynamical model is interpreted as an additional spatial co-ordinate in the thermodynamic system.  
Fig.~\ref{fig:spacetime} illustrates this mapping for the $1d$ Glauber-Ising model, for which the corresponding thermodynamic system is a variant of the $2d$ Ising model.\footnote{%
This $2d$ model is somewhat unusual in that its vertical (space-like) dimension is defined in terms of a lattice while its horizontal (time-like) dimension is continuous.  Nevertheless, it is a bona-fide model that can be analysed by equilibrium statistical mechanics.}

In the general case, we use the same symbol $\bm{\CC}$ to indicate a trajectory of the dynamical model (as in Sec.~\ref{sec:sens}) and  the corresponding configuration of the $(d+1)$-dimensional thermodynamic model.  We define a Boltzmann distribution for the $(d+1)$-dimensional model by assigning probability $P_\tau(\bm{\CC})$ to configuration $\bm{\CC}$.   This means that fluctuations in the dynamical model can (in principle) be analysed by applying methods of equilibrium statistical mechanics to the Boltzmann distribution of the $(d+1)$-dimensional system.

Consider large deviations of some dynamical quantity $\overline{u}$ that corresponds to 
an \emph{intensive} variable in the $(d+1)$-dimensional system.\footnote{%
For our purposes, an extensive variable can be defined (loosely) as a quantity that is obtained by integrating a local quantity over a large system. 
The quantity $\overline{u}$ is intensive if and only if $\tau N \overline{u}$ is extensive.
}
For the example of the Ising model, we consider the time-averaged energy per spin:
\beq
\overline{u}(\bm{\CC}) = \frac{1}{\tau} \int_0^\tau \varepsilon(\CC_t)  \mathrm{d}t \; , 
\qquad \varepsilon(\CC) = \frac{-1}{2N} \sum_{i} \sigma_{i} \sigma_{i+1} \; .
\eeq

For a general dynamical model (with finite $N$) that falls in the scope of Sec.~\ref{sec:scope}, large deviations of $\overline{u}$ can be analysed following Sec.~\ref{sec:theory}.  It is convenient to perform this analysis by setting $\overline{b}_\tau=N\overline{u}$.
Then the biased ensemble of (\ref{equ:s-ens}) is
\beq
P^s_\tau( \bm{\CC} ) =  P_\tau( \bm{\CC} ) \ee^{-s\tau N \overline{u}(\bm{\CC})  - G_N(s,\tau) }
\label{equ:Psu}
\eeq
with
\beq
G_N(s,\tau)=\log \big\langle \ee^{-s \tau N \overline{u}(\bm{\CC})} \big\rangle \; ,
\label{equ:G_N}
\eeq
analogous to (\ref{equ:cgf}).
Recalling that $P_\tau( \bm{\CC} )$ is a Boltzmann distribution for the $(d+1)$-dimensional model, we identify $P^s_\tau( \bm{\CC} )$ in (\ref{equ:Psu}) as a Boltzmann distribution where the energy has been perturbed by the extensive quantity $sN\tau \overline{u}$.\footnote{%
The temperature of the $(d+1)$-dimensional model has been set to unity.}  
Also $G_N$ is the difference in free energy between the perturbed and unperturbed models.  Since $\overline{u}$ was assumed to be intensive, this $(d+1)$-dimensional system has an extensive energy function.  On general thermodynamic grounds~\cite{sokal93} one therefore expects for $N,\tau\to\infty$ that
\beq
\frac{1}{ N \tau } G_N(s,\tau) \to  {\cal G}(s)
\label{equ:GN-thermo}
\eeq 
where ${\cal G}$ is the bulk free-energy density.  We recall from thermodynamics that there are no phase transitions in finite systems: in the present context this means that $G_N(s,\tau)$ should always be an analytic function of $s$.  However, the limiting function ${\cal G}(s)$ may have singularities, which correspond to thermodynamic phase transitions in the $(d+1)$-dimensional model.  In the dynamical context, we refer to these as space-time phase transitions.

\subsection{Space-time phase transitions}
\label{sec:spacetime-dpt}

\begin{figure}
    \includegraphics[width=81mm]{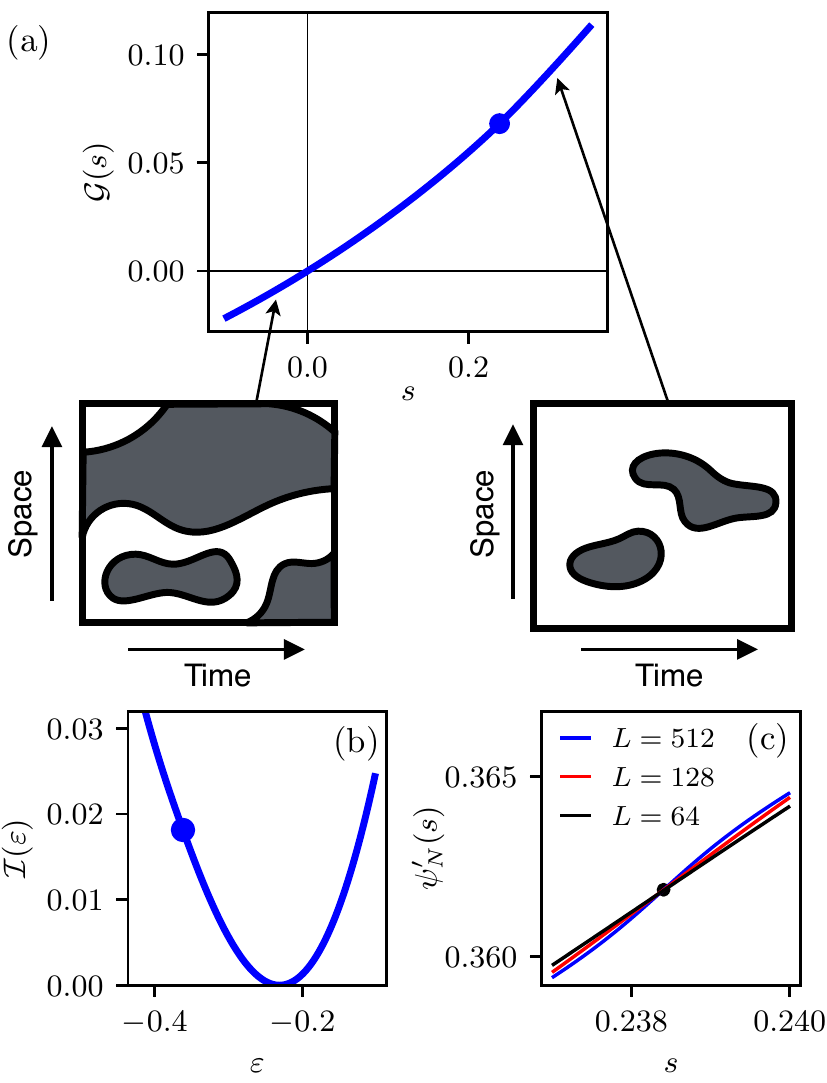}
    \caption{Space-time phase transition for large deviations of the time-integrated energy in the $1d$ Ising model with Glauber dynamics at inverse temperature $
    \beta=1$, using analytical results given in Appendix~\ref{app:ising}, see~\cite{Jack2010}.
    (a) The limiting form of the SCGF as defined in (\ref{equ:GG-lim}).  The filled circle indicates a critical point at $s=s_c$.  Sketches of two representative trajectories are given, recall Fig.~\ref{fig:spacetime}.  For $s>s_c$ then trajectories exhibit long-ranged order in space and time, they resemble ferromagnetic configurations in a $2d$ Ising model.  For $s<s_c$ there is no long-ranged order (in the displayed range), so the behaviour resembles a paramagnet. (b)~The limiting form of the rate function
    as defined in (\ref{equ:cal-I}), the circle indicates the critical point.  (c)~The first derivative of the free energy in systems of three different sizes.}
    \label{fig:ising}
\end{figure}

To analyse these phase transitions, it is convenient to first take $\tau\to\infty$ at fixed $N$, and then later $N\to\infty$.   
At fixed $N$, we define a SCGF by analogy with (\ref{equ:psi})
\beq
\psi_N(s) = \frac1N \lim_{\tau \to\infty} \frac{1}{\tau} G_N(s,\tau)
\label{equ:psi-N}
\eeq
and a rate function by analogy with (\ref{equ:rate-fn})
\beq
I_N(u) = \frac{1}{N} \lim_{\tau\to\infty} \frac{1}{\tau} \log p_N(u|\tau)  \; .
\label{equ:I-N}
\eeq
(The factors of $N$ are included for later convenience.)  The assumptions of Sec.~\ref{sec:scope} are sufficient to ensure that $\psi_N$ and $I_N$ are analytic and strictly convex.  The analogue of (\ref{equ:psi-legendre}) is $\psi_N(s) = \sup_u [ -s u - I_N(u) ] \; .$
For large systems we are motivated by (\ref{equ:GN-thermo}) to define
\beq
 {\cal G}(s) = \lim_{N\to\infty} \psi_N(s)
 \label{equ:GG-lim} 
\eeq
and also
\beq
{\cal I}(b)=\lim_{N\to\infty} I_N(b) \; . 
\label{equ:cal-I}
\eeq 
These functions may not be analytic.
However, the convexity of $I_N$ means that 
\beq
{\cal I}(u) = \sup_s [ -su - {\cal G}(s)  ] \; .
\label{equ:calI-legendre}
\eeq


As an example of a dynamical phase transition, Fig.~\ref{fig:ising} shows the large-deviation behaviour of the time-integrated energy in the $1d$ Glauber-Ising model.  Exact results are available for this model, see~\cite{Jack2010} and also Appendix~\ref{app:ising}.  We show results at inverse temperature $\beta=1$ but the qualitative behaviour is the same for all positive $\beta$~\cite{Jack2010}.  
There is a critical point at $s=s_c$ where $\cal G$ is singular, and there is a corresponding singularity in ${\cal I}$.  This critical point separates a paramagnetic regime for small $s$ and a ferromagnetic regime for $s>s_c$, as might be anticipated by the correspondence with the $2d$ Ising-like model shown in Fig.~\ref{fig:spacetime}.  The transition may also be analysed via a mapping to a quantum phase transition~\cite{Sachdev-book}, see~\cite{Jack2010}.

In finite systems the function $\psi_N(s)$ is analytic, as is $I_N(s)$.  However the second derivative $\psi''_N(s_c)$ diverges logarithmically with $N$: this is the (weak) specific-heat singularity of the $2d$ Ising universality class~\cite{Sachdev-book}.  The singularity is illustrated in Fig.~\ref{fig:ising}(c) by plots of $\psi'_N(s)$ close to $s_c$; its gradient $\psi''_N(s_c)$ grows (slowly) with $N$.

\subsection{First order phase transitions and dynamical phase coexistence}

In thermodynamics, first-order phase transitions are associated with phase coexistence phenomena.  The same situation holds at first-order space-time phase transitions.  However, the manifestation of this phenomenon may differ between thermodynamic and dynamical transitions.  This
can be illustrated by the finite-size scaling behaviour at these transitions~\cite{Jack2006-spacetime,Garrahan2007,Elmatad2010,Nemoto2017first}.
We summarise the associated behaviour, a more detailed analysis can be found in~\cite{Nemoto2017first,Jack2019ising}.   

Applying the thermodynamic analogy of Sec.~\ref{sec:spacetime-dpt}, note that the associated 
thermodynamic model is anisotropic because the horizontal (time-like) and vertical (space-like) axes in Fig.~\ref{fig:ising} are not equivalent.
To reflect this, consider a $d$-dimensional system with $N=L^d$ so that $G_N(s,\tau)$ depends separately on $L$ and $\tau$.
In the analogy with thermodynamics, $L^d\tau$ corresponds to the volume of the thermodynamic model, and $\tau/L$ to its aspect ratio. 

In large deviation analysis, a natural approach is to first take $\tau\to\infty$ at fixed $L$ as in (\ref{equ:psi-N}), and then take $L\to\infty$ as in (\ref{equ:GG-lim}).  This means that the aspect ratio $(\tau/L)\to\infty$.  However, in thermodynamic finite-size scaling analyses, it is more common to consider isotropic systems where the aspect ratio is fixed at unity~\cite{Borgs1990}, this corresponds to taking $L,\tau\to\infty$ together.  Nevertheless, thermodynamic systems with diverging aspect ratio have been analysed~\cite{Privman1983}: they provide a suitable comparison point for large-deviation analyses~\cite{Nemoto2017first,Jack2019ising}.
Limits where $L,\tau\to\infty$ together have also been considered in numerical studies of large deviations~\cite{Hedges2009,Elmatad2010}.

\begin{figure}
    \includegraphics[width=81mm]{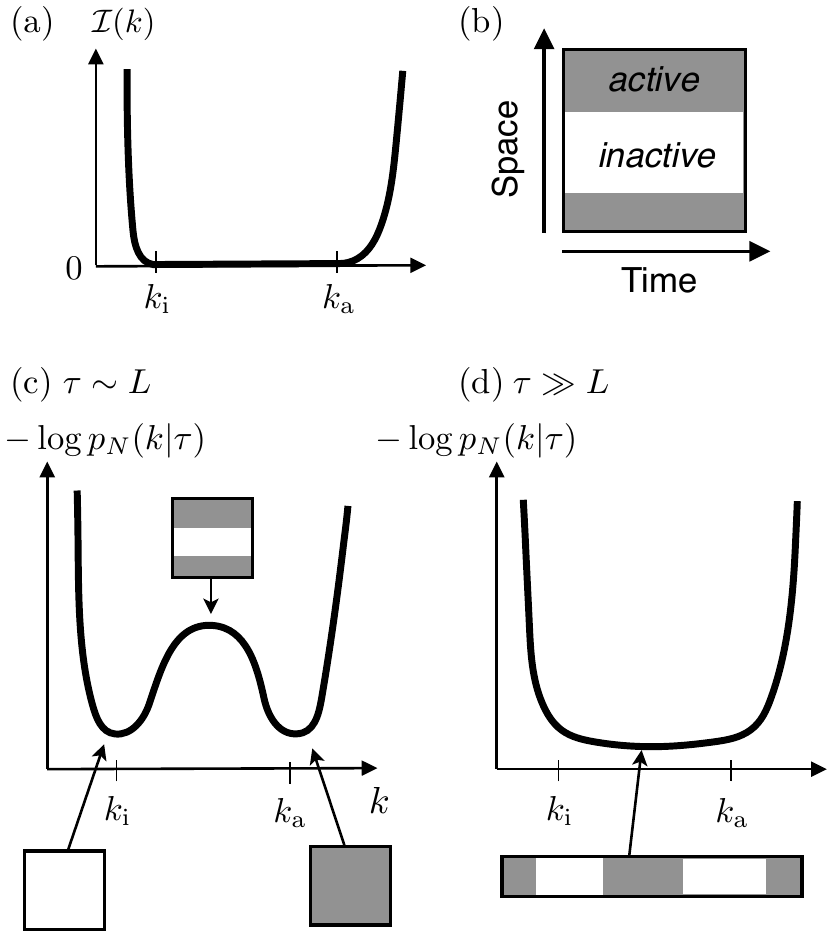}
\caption{Phase coexistence in space-time.  (a)~The limiting rate function defined as in (\ref{equ:cal-I}) for a system exhibiting dynamical phase coexistence between two phases in which the order parameter has values $k_{\rm i}$ and $k_{\rm a}$. (b)~A trajectory exhibiting phase coexistence contains domains of both two dynamical phases, which are labelled as inactive ($k\approx k_{\rm i}$) and active ($k\approx k_{\rm a}$), see for example~\cite{Merolle2005,Jack2006-spacetime}. (c) Sketch of the probability distribution of $k$ in a finite system where $\tau$ is comparable to $L$, with sample trajectories that correspond to different values of $k$.  Dynamical phase coexistence similar to (b) is associated with a local minimum of the probability, which is a local maximum in this plot.  (d) Sketch of the same probability distribution in a finite system where $\tau$ is very large, compared to $L$.  In this case the distribution is unimodal and phase coexistence involves multiple domains arranged along the time-like axis.  See~\cite{Privman1983,Borgs1990} and~\cite{Nemoto2017first,Jack2019ising}}
\label{fig:coex}
\end{figure}

The key fact is that physical behaviour at phase coexistence depends on the aspect ratio of the system. The situation is summarised in Fig.~\ref{fig:coex}.
For $\tau/L=O(1)$ one observes the familiar behaviour of thermodynamic phase coexistence, which means that the probability density $p_N(b|\tau)$ is bimodal with two peaks corresponding to the coexisting phases, see Fig.~\ref{fig:coex}(c) and also~\cite{Hedges2009,Elmatad2010}.  The trough between the peaks corresponds to coexistence, where macroscopic domains of the phases are separated by an interface.  On the other hand, if one takes instead a very large aspect ratio ($\tau\to\infty$ before $L\to\infty$) then $I_N(b|\tau)$ in (\ref{equ:I-N}) is strictly convex so $p_N$ is unimodal.  In this case typical trajectories include many large domains of each phase, which are arranged along the time-like axis, see Fig.~\ref{fig:coex}(d) and also~\cite{Nemoto2017first,Jack2019ising}.

To summarise the central message of space-time thermodynamics: large-deviation theory can be applied to time-averages of (spatially) intensive quantities.  The results can be understood by analogy with $(d+1)$-dimensional thermodynamic systems.  A natural approach to this limit is to consider the behaviour of $G_N$ and $I_N$ as $N\to\infty$, which means that we take a limit of large time \emph{before} any limit of large $N$.  In this case $G_N$ and $I_N$ are both analytic convex functions that converge to non-analytic limits as $N\to\infty$.  This signals that a space-time phase transition is taking place. 

\section{Glassy systems and metastability}
\label{sec:glass}

Interesting examples of space-time phase transitions appear in glassy systems, including supercooled liquids~\cite{Hedges2009}.
The dynamical behaviour of these systems continues to challenge theoretical understanding~\cite{Berthier2011,Chandler2010}.  
The structural relaxation time of a liquid is the time required for a molecule to diffuse a distance comparable with its (microscopic) diameter.
In a simple liquid at a moderate temperature, this time might be a few picoseconds.  
On cooling through the glass transition, the structural relaxation time increases rapidly and 
eventually exceeds the (macroscopic) experimental time scale, which might be seconds or hours.  For practical purposes, the system is no longer ergodic.  The spatial correlations between molecules changes only slightly as the system approaches its glass transition, but the system's dynamical properties change dramatically.

\subsection{Dynamical phase transitions in glasses}
\label{sec:kcm-general}

Observing that the glass transition is a dynamical phenomenon, Merolle, Garrahan and Chandler~\cite{Merolle2005} applied thermodynamic methods to the statistics of $(d+1)$-dimensional trajectories, similarly to Sec.~\ref{sec:spacetime-dpt} above, see also~\cite{Jack2006-spacetime}.  Their idea was that this methodology might capture information that is not available from standard thermodynamic methods.
Early studies~\cite{Merolle2005,Jack2006-spacetime} focussed on simple kinetically-constrained lattice models (KCMs), which capture many of the dynamical features of glassy systems~\cite{Chandler2010}.  
They considered fluctuations of the time-averaged dynamical activity, which in spin models is defined by counting the total number of configuration changes in a trajectory.  This is a proxy for the extent to which molecules in a supercooled liquid are able to move around and explore their environment~\cite{Hedges2009}.

The connection of~\cite{Merolle2005,Jack2006-spacetime} to large deviation theory was realised shortly afterwards, and it was shown that dynamical phase transitions occur generically in KCMs~\cite{Garrahan2007,Garrahan2009}.  This result is discussed in Sec.~\ref{sec:kcm-transition}, below.  It is notable because KCMs do not exhibit thermodynamic phase transitions, raising the possibility that the experimental glass transition might be related to an underlying dynamical phase transition, even in a system with simple thermodynamic properties~\cite{Chandler2010}.  

Following this work on KCMs, numerical studies of atomistic models of liquids have shown evidence for dynamical phase transitions~\cite{Hedges2009,Speck2012-sens,Malins2012-sens,Pitard2011,Fullerton2013,Turci2017prx}.  
Large deviations have been analysed for a variety of time-averaged quantities including several different definitions of dynamical activity~\cite{Pitard2011,Speck2012-sens,Fullerton2013}, and measures of liquid structure~\cite{Malins2012-sens,Turci2017prx}.  There is also evidence for dynamical phase transitions in experiments on glassy colloidal systems~\cite{pinch17,Abou2018}.  Some glassy spin models have thermodynamic glass transitions, and numerical and analytic arguments indicate that these models should also support dynamical transitions~\cite{Jack2010rom}.
Together, these works show that glassy systems generically exhibit large fluctuations, 
which can be probed by a variety of time-averaged quantities, and 
can be characterised via rate functions.

To explain the dynamical phase transition that takes place in KCMs, we discuss the prototypical example of the Fredrickson-Andersen (FA) model~\cite{fa1984} in one dimension.  This was one of the first glassy systems~\cite{Merolle2005} for which large deviations were analysed.  The existence of the phase transition can be proved by a very simple argument~\cite{Garrahan2007,Garrahan2009}.  More recent work has characterised this transition in detail~\cite{Bodineau2012cmp,Bodineau2012jsp,Nemoto2017first,banuls2019}, as well as other large-deviation properties of this model~\cite{Jack2014-east,Nemoto2014-fa,banuls2019}.

\subsection{Dynamical phase transition in the FA model}
\label{sec:kcm-transition}

The FA model (in one dimension) consists of $N$ spins in a linear chain with periodic boundaries.  The state of the $i$th spin is $n_i=0,1$ and a configuration of the system is $\CC=(n_1,n_2,\dots,n_N)$.  Spins with $n_i=1$ are active and indicate excitations, which are regions of a glassy system where particles are moving more than is typical.  Spins with $n_i=0$ are inactive.  The kinetic constraint is that spin $i$ can change its state only if at least one of its neighbours $n_{i\pm 1}$ are active.  If this constraint is satisfied then spin $i$ flips from state $0$ to state $1$ with rate $c$, while the reverse process happens with rate $1-c$.   

\begin{figure}
    \includegraphics[width=85mm]{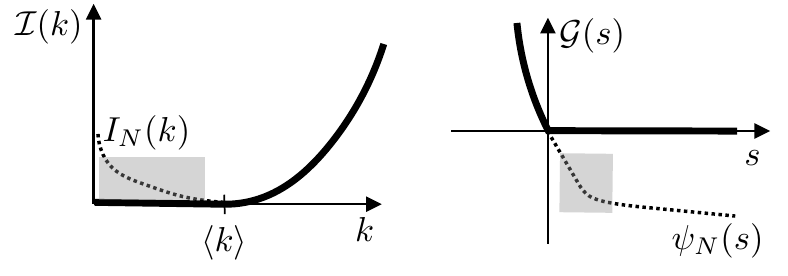}
    \caption{Dynamical phase transition for large deviations of the activity in the FA model.  Heavy solid lines are sketches of the scaled rate function ${\cal I}$ and 
    the bulk free energy density ${\cal G}$.  As  discussed in the  text, ${\cal G}$ is singular at $s=0$ and ${\cal G}(s)=0$ for $s\geq0$.  Similarly ${\cal I}(k)=0$ for all $k\leq\langle k\rangle$.  Dotted lines indicate the qualitative behaviour in finite systems, they are sketches of $I_N(k)$
    and $\psi_N(s)$ as defined in (\ref{equ:I-N},\ref{equ:psi-N}).  These functions are convex and analytic but they converge to the (non-analytic) limits ${\cal I}$, ${\cal G}$ as $N\to\infty$.  The shaded boxes indicate the range over which the  finite-size scaling analysis of~\cite{Nemoto2017first} is relevant.   
    }
    \label{fig:kcm}
\end{figure}

The behaviour of the model depends on the parameter $c$.  In particular, for a system at equilibrium then the fraction of spins that are in state $1$ is $\langle n_i\rangle = c$.  The dependence of the model on temperature $T$ is captured by identifying $c=\ee^{-J/(\kB T)}$ where $J$ is the characteristic energy of an active site (excitation).\footnote{%
Note also: if $n_i=0$ for all $i$ then the configuration of the system can never change.  For studies of large deviations it is therefore convenient to define the model on a configuration space that excludes this configuration~\cite{Garrahan2009}.  In this case the model is irreducible and falls within the scope of Sec.~\ref{sec:scope}.
}%

Now let $k_{i,\tau}$ be the number of times that spin $i$ changes its state, between time zero and time $\tau$.  Summing over all spins, a time-averaged (intensive) measure of dynamical activity is 
\beq
\overline{k}_\tau = \frac{1}{N\tau} \sum_i k_{i,\tau} \;.
\eeq  
This corresponds to (\ref{equ:bbar-alpha}) with $\alpha(\CC,\CC')=(1/N)$ for all $\CC,\CC'$.  

We analyse the large deviations of this activity by following Sec.~\ref{sec:spacetime-dpt} with $\overline{u}\to\overline{k}_\tau$ (this is similar to Sec.~\ref{sec:def}, replacing $\overline{b}_\tau\to N\overline{k}_\tau$).   The following very simple argument shows that the functions ${\cal G}$ and ${\cal I}$ have singularities that correspond to first-order phase transitions.   Consider the configuration with $n_1=1$ and $n_i=0$ for all other sites.  The rate of transitions out of this configuration is $2c$; the probability that it occurs as initial condition is denoted by $\pi_1$.

Now define a very simplistic controlled process where the system begins in this configuration and never leaves it.  For this trajectory one has 
\beq
\frac{ P_\tau^{\rm con}(\bm{\CC}) }{ P_\tau(\bm{\CC}) } = \frac{1}{ \pi_1 \ee^{-2c\tau} }
\label{equ:Pcon-fa}
\eeq
Using this result with (\ref{equ:G-var},\ref{equ:GN-thermo},\ref{equ:GG-lim}) and noting that $\pi_1$ is independent of $\tau$, one obtains
$\psi_N>-2c/N$ and hence
\beq
{\cal G}(s) \geq 0 \; .
\label{equ:Gs-zero-kcm}
\eeq
Since the activity $\overline{k}_\tau\geq0$, it follows from (\ref{equ:cgf},\ref{equ:Gs-zero-kcm}) that ${\cal G}(s)=0$ for $s\geq0$.

Fig.~\ref{fig:kcm} illustrates the result: there is a discontinuity in the first derivative of ${\cal G}$ at $s=0$, which corresponds to a first-order space-time phase transition.\footnote{%
In order to establish this one must show that $\lim_{s\uparrow0}{\cal G}'(s)>0$, this is straightforward~\cite{Garrahan2009}.
}
 Applying (\ref{equ:I-legendre}), it follows that ${\cal I}(k)=0$ for all $k<\langle \overline{k}_\tau\rangle$.   This means that for large $N,\tau$, rare events where $\overline{k}_\tau$ is smaller than its average have log-probabilities that do not scale as $N\tau$.  In fact, these log-probabilities are much smaller: they are either proportional to $N$ or $\tau$, depending on the relative magnitudes of these two quantities~\cite{Jack2006-spacetime}.  

We note that the bound (\ref{equ:Gs-zero-kcm}) is very general in KCMs, and establishes that these phase transitions occur in many different models~\cite{Garrahan2007,Garrahan2009}.  
However, it does rely on the existence of a ``hard'' kinetic constraint, which means that for a typical configuration $\CC$, there are spins which cannot flip.
This is a strong assumption and leaves open the question as to whether similar phase transitions are possible in models with softened constraints as in~\cite{Elmatad2010}, where every spin flips with a non-zero rate.  
In fact similar (first-order) dynamical phase transitions still occur in the softened FA model~\cite{Elmatad2010}, although in this case the singularity in ${\cal G}(s)$ occurs at $s^*>0$, and the only zero of ${\cal I}(k)$ is at $k=\langle \overline{k}_\tau\rangle$.

\subsection{Large deviations and metastable states}

These results for kinetically constrained models show that glassy systems with simple thermodynamic properties can still exhibit dynamical phase transitions.  
However, other theories of the glass transition assert that slow relaxation in liquids is linked to long-lived metastable states that can be analysed thermodynamically.  This theoretical paradigm is certainly valid in a class of mean-field spin glasses,\footnote{%
In this context, the ``mean-field'' nomenclature means that the strength of the interaction between spins is independent of the distance between them.}
while research continues into the question of whether it applies in physical (three-dimensional) liquids~\cite{berthierEdiger2016}.
Some mean-field spin-glass models exhibit first-order dynamical phase transitions~\cite{Jack2010rom}, similar to those in kinetically constrained models.  The operator approach of Sec.~\ref{sec:spectral} has been used to show that long-lived metastable states lead naturally to such transitions~\cite{Jack2010rom}. Here we give a brief explanation as to how the same conclusions can be reached (perhaps more intuitively) by an optimal-control argument.

Metastability is associated with a separation of time scales.  The physical idea  -- which can be applied in non-equilibrium systems as well as in equilibrium~\cite{Gaveau1996,Gaveau1998,Biroli2001} -- is that if a system is initialised in a metastable state then it equilibrates quickly within that state, on a time scale $\tau^{\rm f}=O(1)$, before eventually relaxing to some other state on a much longer time scale $\tau^{\rm s}\gg 1$.  

Consider a system with $n\geq 2$ states, labelled by $\alpha=1,2,\dots,n$.  This includes the case where one state is stable and the others are metastable (for example a mean-field ferromagnet in a field).  It also includes systems at thermodynamic phase coexistence, which have two or more stable states.\footnote{%
For the purposes of this discussion, the difference between stable and metastable states is that metastable states have a vanishing probability in the steady state.}
Let $\pi_\alpha$ be the probability that a steady-state configuration belongs to state $\alpha$.
We analyse large deviations of an intensive observable $\overline{u}$ that has different average values in each state: we denote these averages by $\langle u\rangle_\alpha$.  In cases where the time scales are well-separated and the metastable states are well-defined then $\sum_\alpha \pi_\alpha\simeq 1$ and the steady-state average of $u$ is $\langle u\rangle \simeq \sum_\alpha \pi_\alpha \langle u\rangle_\alpha$.  These approximate equalities are accurate if $\tau_{\rm s}\gg \tau_{\rm f}$.

Following (\ref{equ:Pcon-fa}) as well as~\cite{Jack2010rom,Jack2011} we consider a controlled process that starts in state $\alpha$ and remains there for the entire trajectory.  Its behaviour within state $\alpha$ matches the natural dynamics of the model within that state. Since relaxation is fast within the metastable state, the time for the original (uncontrolled) model to leave this state is exponentially distributed with a mean that we denote by $\tau_\alpha^{\rm s}$.  By analogy with (\ref{equ:Pcon-fa}), we deduce that 
\beq
\frac{ P_\tau^{\rm con}(\bm{\CC}) }{ P_\tau(\bm{\CC}) } \simeq \frac{1}{ \pi_\alpha  \ee^{-\tau/\tau_\alpha^{\rm s}}  } \; .
\label{equ:Pcon-alpha}
\eeq
 
As usual we consider large systems, $N\to\infty$.   In idealised cases such as mean-field ferromagnets, the slow relaxation between states occurs on time scale $\tau^{\rm s}\sim \ee^{\kappa N}$ where $N$ is the system size and $\kappa=O(1)$.   If state $\alpha$ is metastable then $\pi_\alpha \sim \ee^{-N\Delta f}$ where $\Delta f=O(1)$ is a difference in (intensive) free energy; if $\alpha$ is stable then $\pi_\alpha=O(1)$. 
 Using (\ref{equ:Pcon-alpha}) with (\ref{equ:G-var}) and $\overline{b}=N\overline{u}$ shows that 
 \beq
 G_N(s,\tau) \geq - sN\tau\langle u\rangle_\alpha  + \log\pi_\alpha - (\tau/\tau_\alpha^{\rm s}) \; .
 \eeq
Taking $\tau\to\infty$ at fixed $N$ and using (\ref{equ:psi-N},\ref{equ:GG-lim}) gives 
\beq
{\cal G}(s) \geq -s\langle u\rangle_\alpha - \lim_{N\to\infty} (N\tau_\alpha^{\rm s})^{-1}  \; .
\label{equ:GG-meta}
\eeq
If $\tau_\alpha^{\rm s}\gg1$ is a slow time scale then one sees that ${\cal G}(s) \geq -s\langle u\rangle_\alpha$.  Using also that $\psi_N'(0)=-\langle u \rangle$
and $\langle u \rangle\neq \langle u \rangle_\alpha$, this implies that ${\cal G}'(s)$ has a discontinuity at $s=0$, which corresponds to a first-order space-time phase transition, similar to the case of kinetically constrained models.  A more detailed analysis of this case can be found in~\cite{Jack2010rom}, using the operator approach.

We emphasise that such first-order transitions are generic for systems where $(\tau^{\rm s}/\tau^{\rm f})\to\infty$ which includes mean-field systems with metastable states, and finite-dimensional systems at phase coexistence.  For finite-dimensional systems away from phase coexistence then all metastable states have finite lifetimes, and one expects $(\tau_\alpha^{\rm s})^{-1}\propto N$.  (For example, recall that nucleation rates for systems close to phase coexistence are proportional to the system size $N$~\cite{Auer2001}.)  In such cases, (\ref{equ:GG-meta}) gives a bound on ${\cal G}$ that is not sufficient to establish the existence of a phase transition, but can be used to relate crossovers in ${\cal G}(s)$ and $G_N(s,\tau)$ to properties of metastable states, particularly $\langle u \rangle_\alpha$ and $\tau^{\rm s}_\alpha$~\cite{Jack2010rom,Jack2011}.   These arguments establish strong connections between metastability and large deviations, which (we argue) are very useful when interpreting large-deviation computations for glassy systems~\cite{Hedges2009,Malins2012-sens,Turci2017prx}.
 
\section{Fluctuation theorems and time's arrow}
\label{sec:fluct}

Glassy systems have slow dynamics but their equilibrium states are time-reversal symmetric.  We now turn to models of non-equilibrium steady states.   
Early work in this area~\cite{Gallavotti1995,Lebowitz1999} demonstrated the usefulness of large deviation studies of time-averaged quantities in physics,
by exploiting connections between dissipation and irreversibility.
In this section we set Boltzmann's constant $\kB=1$, so that entropy is a dimensionless quantity.

We write $\bm{\CC}^{\rm R}$ for the trajectory that is obtained by reversing the arrow of time in trajectory $\bm{\CC}$.  In the simplest case, this means that $\CC^{\rm R}_{\tau-t} = \CC_t$.  More generally the time-reversal operation might involve a change in some system variables, such as reversal of molecular velocities, as in~\cite{Crooks1998,Kaiser2017ent}.  Then, a (time-integrated) measure of irreversibility for trajectory $\bm{\CC}$ within a given model can be identified as 
\beq
\Sigma_\tau(\bm{\CC}) = \log \frac{ P_\tau(\bm{\CC}) }{ P_\tau(\bm{\CC}^{\rm R}) }
\label{equ:Sigma-v1}
\eeq
Recall that $P$ includes the probability of the initial configuration of the system $\CC_0$ and that these initial conditions are taken from the steady state 
of the system. It follows that for equilibrium systems, $\Sigma_\tau(\bm{\CC})=0$ exactly, for every time $\tau$ and every trajectory $\bm{\CC}$.

It is useful to define 
\beq
P^*_\tau(\bm{\CC})=P_\tau(\bm{\CC}^{\rm R})
\label{equ:Padj}
\eeq
 and to identify this $P^*$ as the probability distribution for trajectories under a particular controlled process which we refer to as the adjoint process, following~\cite{Bertini2015}.  One sees that 
\beq
\langle \Sigma_\tau(\bm{\CC}) \rangle = {\cal D}(P || P^*) \geq 0 \; .
\eeq
The mean entropy production can never be negative; it is zero only for time-reversal symmetric (equilibrium) systems, since $P=P^*$ in that case.

One drawback of the irreversibility measure $\Sigma_\tau$ is that the quantity $P_\tau(\bm{\CC})$ appearing in (\ref{equ:Sigma-v1}) cannot usually be evaluated, because it depends on the probability of  the initial state  of the trajectory, which is typically not known (except in equilibrium systems where it only depends on the energy).  However, one may define a time-averaged rate of entropy production as
\beq
\overline{\sigma}_\tau = \frac{1}{\tau}  \left[ \Sigma_\tau(\bm{\CC}) - \log \frac{\pi(\CC_0)}{\pi(\CC^{\rm R}_0)} \right]
\label{equ:sigma}
\eeq
where we recall that $\pi(\CC_0)$ is the probability density for the initial condition of the trajectory.\footnote{
Note $\CC^{\rm R}_0$ is the initial condition of the time-reversed trajectory which in the simplest case coincides with $\CC_\tau$.}  
In non-equilibrium systems, the usual situation is that $\Sigma_\tau$ grows with $\tau$ while $\log \frac{\pi(\CC_0)}{\pi(\CC^{\rm R}_0)}$ remains finite.\footnote{
This is certainly the case if $\pi(\CC)$ is bounded, which holds for the finite systems of Sec.~\ref{sec:scope}.
}  
In this case, the large deviations of $\overline{\sigma}_\tau$ are the same as the large deviations of $\Sigma_\tau/\tau$, even if these quantities have different values when $\tau$ is finite.

\begin{figure}
   \includegraphics[width=80mm]{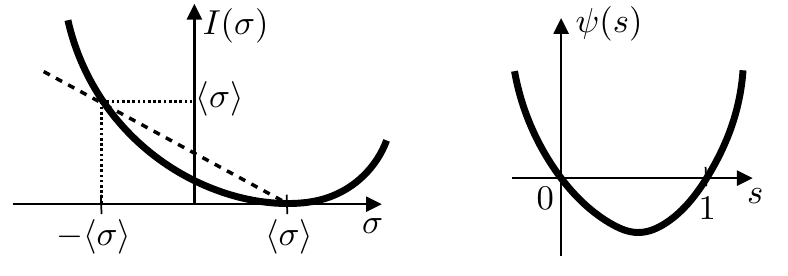}
    \caption{Sketch of the rate function and SCGF for the entropy production in a generic system  that obeys the fluctuation theorem.  The dashed line in the left panel has gradient $-1/2$, it intersects the rate function at $\sigma=\pm\langle\sigma\rangle$, consistent with (\ref{equ:fluct-I}).  The SCGF has reflection symmetry through $s=\frac12$, consistent with (\ref{equ:psi-gallavotti}).}
    \label{fig:entropy-prod}
\end{figure}

In many physical systems, closed formulae for $\overline{\sigma}_\tau$ are available.  For example, consider a simple model for particle motion (in $d$ dimensions)
\beq
\mathrm{d}{\CC}_t = \frac{f(\CC_t) }{\gamma} \mathrm{d}t + \sqrt{\frac{2T}{\gamma} } \mathrm{d}W_t \; ,
\label{equ:sde-simple-gamma}
\eeq
similar to (\ref{equ:sde-simple}).
The natural physical interpretation of this model is that a particle moves through a viscous fluid with friction constant $\gamma$ at temperature $T$, and feels an non-conservative external force $f(\CC_t)$. 
Then it may be shown from  (\ref{equ:sigma}) that
\beq
\overline{\sigma}_\tau = \frac{1}{\tau T} \int_0^\tau f(\CC_t) \circ \mathrm{d}\CC_t \; .
\label{equ:sigma-sde}
\eeq
We identify $\tau T \overline{\sigma}_\tau$ as the total work done by the force $f$ which coincides (in this simple situation) with the heat dissipated in the fluid.  Dividing the dissipated heat by the temperature gives the entropy production, so the probabilistic definition of $\overline{\sigma}_\tau$ in  (\ref{equ:sigma})  coincides with the time-averaged rate of (physical) entropy production.  

Note that heat and work coincide in this example system because all forces were assumed to be external: 
hence there is no internal energy.  
To separate the definitions of heat and work one should formulate the first law of thermodynamics by defining an internal energy $U$ and a corresponding force $-\nabla U$.  Then write $f=F^{\rm ext}-\nabla U$ in (\ref{equ:sde-simple-gamma}), where $F^{\rm ext}$ is an external force~\cite{Seifert2012review}.  The work is then $\int F^{\rm ext}(\CC_t) \circ \mathrm{d}\CC_t$, and the heat transferred to the fluid is 
$\int_0^\tau f(\CC_t) \circ \mathrm{d}\CC_t$, consistent with (\ref{equ:sigma}).  The difference of these quantities is the change in internal energy: this is the first law of thermodynamics.

For Markov chains with jump rates $W(\CC\to\CC')$, the analogue of (\ref{equ:sigma-sde}) is given by (\ref{equ:bbar-alpha}) with $\alpha(\CC,\CC') = \log [W(\CC\to\CC')/W(\CC'\to\CC)]$, which requires the assumption that $W(\CC\to\CC')$ is non-zero whenever $W(\CC\to\CC')$ is non-zero.  (This property is sometimes called weak reversibility.)

Returning to the main argument, it follows from the explicit formula (\ref{equ:sigma}) that large deviations of $\overline{\sigma}_\tau$ can be analysed within the class of models discussed in Sec.~\ref{sec:scope}.  The connection of the entropy production $\overline{\sigma}_\tau$ with the irreversibility measure $\Sigma_\tau$ means that large deviations of $\overline{\sigma}_\tau$ have interesting symmetry properties, as we now discuss.

\subsection{Fluctuation theorem of Gallavotti-Cohen}

We discuss fluctuation theorems for the entropy production in non-equilibrium steady states~\cite{Gallavotti1995,Evans1993,Sekimoto1998,Lebowitz1999,Maes1999,Crooks2000,Seifert2012review}.
%
Consider first the CGF for $\Sigma_\tau$:
\begin{align}
G^\Sigma(s,\tau) &  = \log \langle \ee^{-s\Sigma_\tau} \rangle
\nonumber \\
& = \log  \int \left( \frac{ P_\tau(\bm{\CC}^{\rm R}) }{ P_\tau(\bm{\CC}) } \right)^{s} P_\tau(\bm{\CC}) \mathrm{d}\bm{\CC}
\nonumber \\
& = \log  \int \left( \frac{ P_\tau(\bm{\CC}) }{ P_\tau(\bm{\CC}^{\rm R}) } \right)^{1-s} P_\tau(\bm{\CC}^{\rm R}) \mathrm{d}\bm{\CC}
\end{align}
where the first line is the definition of $G^\Sigma$, the second line uses (\ref{equ:Sigma-v1}), and the third simply rearranges various terms.
Changing integration variable from $\bm{\CC}^{\rm R}$ to $\bm{\CC}$, and using again (\ref{equ:Sigma-v1}) one finds the symmetry relation
\beq
G^\Sigma(s,\tau) = G^\Sigma(1-s,\tau) \; .
\label{equ:G-gallavotti}
\eeq
See for example~\cite{Lebowitz1999}, where the quantity $\Sigma_\tau$ was denoted by $\overline{W}$.

Now consider large deviations of the entropy production $\overline{\sigma}_\tau$ whose SCGF is denoted here by $\psi(s)$.  
Recalling (\ref{equ:sigma}), one may expect that the large deviations of $\overline{\sigma}_\tau$ are the same as those of $\Sigma_\tau/\tau$, in which case one would have  $\psi(s) = \lim_{\tau\to\infty} \tau^{-1} G^\Sigma(s,\tau)$.  The relationship between $\overline{\sigma}_\tau$ and $\Sigma_\tau$ is discussed in~\cite{Lebowitz1999}, which showed (for several broad classes of stochastic model) that
\beq
\psi(s) = \psi(1-s) \; .
\label{equ:psi-gallavotti}
\eeq
This is the symmetry that was identified by Gallavotti and Cohen~\cite{Gallavotti1995}.  It is closely related to (\ref{equ:G-gallavotti}) but we note that (\ref{equ:psi-gallavotti}) is a statement
about large deviations of $\overline{\sigma}_\tau$ as $\tau\to\infty$, in contrast to (\ref{equ:G-gallavotti}) which is a statement about $\Sigma_\tau$ that is valid for all $\tau$.  See also~\cite{Crooks2000,Seifert2012review}.  

Now assume convexity of $\psi$ and use (\ref{equ:I-legendre}) with (\ref{equ:psi-gallavotti}) to write 
\begin{align}
I(-\sigma) & =  \sup_s [ s\sigma - \psi(1-s) ] 
\end{align}
Relabelling the dummy variable $s=1-x$ and using (\ref{equ:I-legendre}) one obtains a fluctuation theorem~\cite{Lebowitz1999,Maes1999}
\beq
I(-\sigma) = I(\sigma) + \sigma \; .
\label{equ:fluct-I}
\eeq
Taking $\sigma>0$, one sees that $I(-\sigma)$ determines the log-probability of trajectories with negative entropy production.
Since $I(-\sigma)>I(\sigma)$, these 
trajectories are exponentially rarer than trajectories with positive entropy production.  (This can be interpreted as a statistical form of 
the second law of thermodynamics.)
In addition, the difference in log-probability is given 
quantitatively by (\ref{equ:fluct-I}), so the fluctuation theorem (which is an equality) contains more information than the second law (which is an inequality).


For equilibrium systems we recall  that $\Sigma_\tau=0$ exactly, so the methods of large-deviation theory are not relevant.  At a formal level then $I(\sigma)=\infty$  whenever $\sigma\neq0$, and $\psi(s)=0$ for all $s$.   For non-equilibrium systems, it is notable that the optimally-controlled process at $s=\frac12$ is time-reversal symmetric; also the optimally-controlled process at $s=1$ is the adjoint process, which corresponds to the original process running backwards in time, see for example~\cite{Bonanca2016}.

\begin{figure}
   \includegraphics[width=80mm]{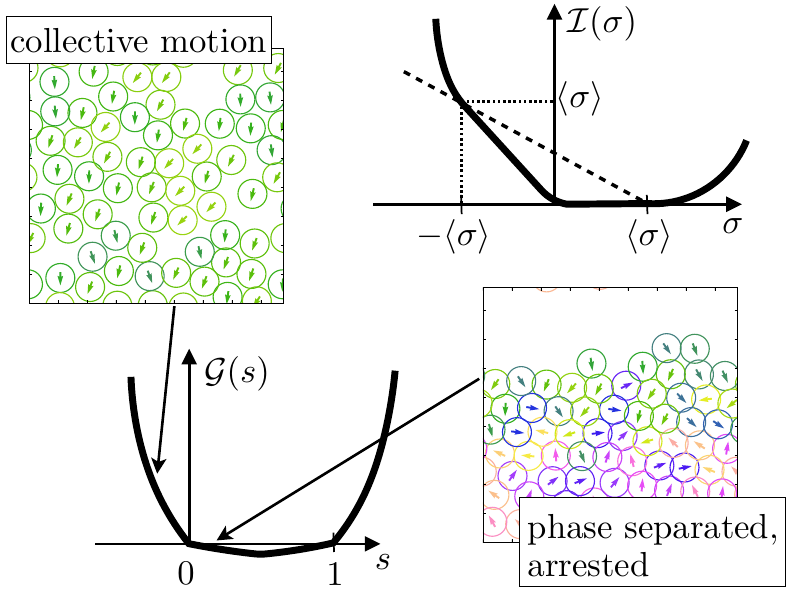}
    \caption{Illustration of the behaviour of the entropy production $\sigma$ in a system of active Brownian particles, based on~\cite{Nemoto2019}.
    We show the limiting rate function ${\cal I}$ and the corresponding free energy ${\cal G}$, which obey a Gallavotti-Cohen symmetry as in Fig.~\ref{fig:entropy-prod}.  The free energy ${\cal G}$ is singular at $s=0$, and ${\cal I}$ has two linear segments, see the text for a discussion.  For $s<0$ (high entropy production), typical configurations involve spontaenous particle alignment and exhibit collective motion. For $0<s<1$ (low entropy production)  system enters a dynamically-arrested phase-separated state.}
    \label{fig:active}
\end{figure}

\subsection{Example: Active Brownian particles}

Fluctuation theorems such as (\ref{equ:psi-gallavotti}) are very general results.  However, the analysis of the fluctuations of the entropy production in specific systems can reveal additional rich structure.  An interesting example is the behaviour of active-matter systems~\cite{Nemoto2019,Cagnetta2017,Grand2018,Fodor2020}.  As an example we consider a system of active Brownian particles~\cite{Fily2012,Redner2013}, as considered in~\cite{Nemoto2019}.  It consists of $N$ circular particles in a two-dimensional system of size $L^2$.  Particle $i$ has an orientation, which is represented by a unit vector $\bm{e}_i$.  The particles interact by repulsive forces and they undergo thermal diffusion with diffusion constant $D_0$.  In addition, they feel non-conservative propulsive forces of fixed strength which act along their orientation vectors.  The propulsive forces are such that a single isolated particle moves with average speed $v_0$.  Each orientation vector undergoes rotational diffusion, independent of all other co-ordinates.   

Ref.~\cite{Nemoto2019} considered large deviations of a quantity called the active work, which has a corresponding (intensive) measure of entropy production:
\beq
\overline{\sigma}_\tau = \frac{1}{N\tau} \sum_{i=1}^N \int_0^\tau \frac{v_0}{D_0} \bm{e}_i(t) \circ \mathrm{d}{\bm{r}}_i(t) 
\label{equ:k-work-abp}
\eeq  
where ${\bm{r}}_i$ is the position of particle $i$, and the integral is evaluated using the Stratonovich convention.  The physical interpretation of $\overline{\sigma}_\tau$ is that there is a force on particle $i$, acting in direction $\bm{e}_i$  (with constant magnitude).  The integral in (\ref{equ:k-work-abp}) is the work done by this force, normalised by the temperature.
We refer to $\overline{\sigma}_\tau$ as the entropy production, although other definitions of the entropy production are possible in such systems~\cite{Fodor2016,Pietzonka2017-active,Mandal2017,Shankar2018}.  Large deviations of $\overline{\sigma}_\tau$  obey the fluctuation theorems (\ref{equ:psi-gallavotti},\ref{equ:fluct-I}). 

The resulting large-deviation phenomenology is illustrated in Fig.~\ref{fig:active}, following~\cite{Nemoto2019}.  We focus on large systems and we consider ${\cal I}$ and ${\cal G}$ as defined in (\ref{equ:GG-lim},\ref{equ:cal-I}).  Note that ${\cal I}$ obeys the fluctuation theorem (\ref{equ:fluct-I}) but its behaviour is quite different from the illustration in Fig.~\ref{fig:entropy-prod}.  The reason is that this system exhibits several space-time phase transitions which appear in the limit of large system size.  This corresponds to $N\to\infty$ at a fixed overall density $\rho_0 = N/L^2$.  In the following, it is sufficient to consider only $\sigma>0$: the behaviour for $\sigma<0$ follows from the fluctuation theorem.

A first observation is that the behaviour for $\sigma>0$ and $s<\frac12$ in Fig.~\ref{fig:active} somewhat resembles Fig.~\ref{fig:kcm}: there is a discontinuity in ${\cal G}'(s)$ at $s=0$ and a range of $\sigma$ over which ${\cal I}(\sigma)=0$.
  This was explained in~\cite{Nemoto2019} by an optimal-control argument: they proposed a controlled process that can be used with (\ref{equ:Ib-KL}) to show that $I_N(\sigma)\leq O(1/L)$ for a finite range of $\sigma$ between $0$ and $\langle\sigma\rangle$.\footnote{%
An open question from that work is whether this range extends down to $\sigma=0$ or whether there it has a non-zero lower limit.}  
Hence ${\cal I}(\sigma)=0$ in this regime, by (\ref{equ:cal-I}).  The behaviour of the controlled system in this case is that the particles form a high-density cluster where particle motion is strongly reduced, and $\sigma$ is small.   Hence this state was called ``phase-separated and arrested''.  The associated reduction in particle motion is analogous to the transition to the inactive phase in Fig.~\ref{fig:kcm}, which explains the similarity to that case, see also Sec.~\ref{sec:exc}.

For large deviations with $\sigma>\langle\sigma\rangle$, the numerical results of~\cite{Nemoto2019} show spontaneous symmetry breaking, in that particles align their orientations with each other (Fig.~\ref{fig:active}).  In this case they also move collectively through the system.  For an intuitive understanding of this transition, it is useful to consider a controlled system where the particles' orientation vectors feel forces (or torques) that tend to align them.  Ref.~\cite{Nemoto2019} considered a mean-field (infinite-ranged) interaction.  If this interaction is strong enough to create long-ranged (ferromagnetic) order of the orientations, it clearly reduces the number of interparticle collisions, and this increases $\sigma$.  This controlled process provides a bound on ${\cal I}$ via (\ref{equ:Ib-KL}) and numerical tests indicate that this bound is close to the true value of ${\cal I}$.  The conclusion is that particle alignment is an effective mechanism for fluctuations of the entropy production.  

The understanding of large deviations in this system is not yet complete, but it is clear from~\cite{Nemoto2019} that fluctations with $\sigma<\langle\sigma\rangle$ are strongly coupled to density fluctuations and the arrest of particle motion, while fluctuations with $\sigma>\langle\sigma\rangle$ are associated with spontaneous symmetry breaking and particle alignment.  This illustrates the rich large-deviation behaviour of these non-equilibrium systems.

\section{Exclusion processes and hydrodynamic behaviour}
\label{sec:exc}

A very active area of large-deviation research is the behaviour of interacting-particle systems
including exclusion processes and zero-range processes~\cite{Derrida1998,Derrida2007,Bertini2002,Bodineau2004,Bertini2015,Harris2005,Hurtado2014,Baek2017,Tizon2017}.  This Section gives a brief overview of some of the relevant phenomena,  focussing on the similarities and differences between these systems and those analysed in previous Sections.

\subsection{Activity fluctuations in the simple symmetric exclusion process}

\begin{figure}
    \includegraphics[width=85mm]{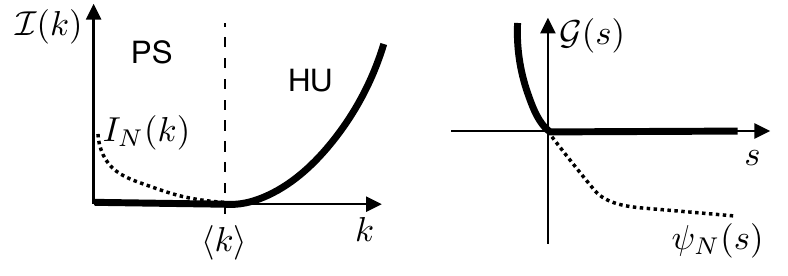}
    \caption{Large deviations of the activity in the SSEP with periodic boundaries.  There are some similarities with the results of Fig.~\ref{fig:kcm} for the FA model: ${\cal G}$ is singular at $s=0$ and ${\cal G}(s)=0$ for $s\geq0$.  Also ${\cal I}(k)=0$ for all $k\leq\langle k\rangle$.  
    Dotted lines indicate the qualitative behaviour in finite systems.  In contrast to the FA model, $\psi''_N(0)$ diverges with $N$ and ${\cal G}''(s)$ diverges as $s\to0^-$.  This is due to slow hydrodynamic fluctuations.  As explained in the text, the system is macroscopically inhomogeneous (phase-separated, PS) for fluctuations with $\overline{k}_\tau<\langle k\rangle$ [$s>0$] and hyperuniform (HU) for  fluctuations with $\overline{k}_\tau>\langle k\rangle$  [$s<0$].  The regime with $s=O(L^{-2})$ can be characterised by macroscopic fluctuation theory, it includes the whole PS regime and corresponds to ${\cal I}(k)=O(1/L)$. }
    \label{fig:ssep}
\end{figure}

As a concrete example, we focus on the  symmetric simple exclusion process (SSEP) with periodic boundaries, as considered in~\cite{Appert2008}
as well as~\cite{Lecomte2012,Jack2015,Brewer2018}. In this case, $N$ particles move on the $L$ sites of a one-dimensional periodic lattice, with at most one particle per site.  Suppose that the particle hop rate is $\gamma$ and the lattice spacing is $a_0$ so that the diffusion constant for a single particle is $D_0=\gamma a_0^2/2$.\footnote{%
In the literature, one often measures time in units where $D_0=1$ but we retain $D_0$ here as a parameter.}   
In this system, exact results are available, for large deviations of the (time-averaged, intensive) particle current $\overline{\jmath}_\tau$ and the activity $\overline{k}_\tau$~\cite{Appert2008}.   The current is defined using (\ref{equ:bbar-alpha}) with $\alpha=1/L$ when a particle hops to the right and $\alpha=-1/L$ for hops to the left.  Similarly the activity is defined by taking $\alpha=1/L$ for all hops.\footnote{%
With this definition, the current and activity are intensive in the sense that $\langle \overline{\jmath}_\tau\rangle=O(1)$ as $N\to\infty$.
In the literature it is more common to work with corresponding extensive observables, but we use the intensive versions here, to facilitate comparison with earlier Sections.
}  
We focus here on large deviations of the activity.

This process is a finite Markov chain, satisfying the conditions of Sec.~\ref{sec:scope}.  
Hence the rate functions for finite systems are analytic and convex.
The interesting behaviour occurs in the limit of large system size, which means that $L\to\infty$ with a fixed mean density $\overline\rho=N/L$. 
This suggests that the space-time thermodynamic theory of Sec.~\ref{sec:spacetime-general} should be applicable.  
However, the number of particles is a conserved quantity in the SSEP, which means that the corresponding $(d+1)$-dimensional thermodynamic model has some unusual features, from the thermodynamic perspective.  

To understand dynamical large deviations, note first that if all the particles in the SSEP form a single cluster by occupying adjacent sites, then there are only two particle hops that are possible (at the edges of the cluster).  In this case one may apply exactly the same argument as Sec.~\ref{sec:kcm-transition} to obtain 
\beq
\psi_N(s) \geq -2\gamma/N
\eeq
where $\psi_N$ is the SCGF [as in (\ref{equ:psi-N})] for the activity $\overline{k}_\tau$.  
The activity $\overline{k}_\tau\geq0$ so $\psi_N(s)\leq0$ for positive $s$. 
 This establishes that $\psi_N(s) = O(1/N)$ for positive $s$ (low activity).  On the other hand $\psi_N(s)$ is of order unity for negative $s$ (high activity).  Defining ${\cal G}$ as in (\ref{equ:GG-lim}) one arrives at a situation similar to Fig.~\ref{fig:kcm}, with ${\cal G}(s)=0$ for $s\geq0$ while ${\cal G}$ is of order unity for $s<0$.  
 
 This result is correct but it misses some important properties of exclusion processes, for which one requires a more detailed analysis~\cite{Appert2008,Lecomte2012}.
The SSEP has a slow diffusive time scale associated with large-scale density fluctuations $\tau_L \sim L^2/D_0$.  These slow (hydrodynamic) fluctuations hinder ergodicity and tend to enhance the variance of time-averaged quantities.  For example, it may be verified from~\cite{Appert2008} that the variance of $\overline{k}_\tau$ behaves for large $N,\tau$ as
$
{\rm Var}(\overline{k}_\tau) \propto 1/\tau
$,
independent of $N$, and hence $\psi_N''(0)=O(N)$.  This is in contrast to dynamical phase coexistence as it occurs in the FA model, where $\psi''_N(0)$ is of order unity. 
%

The activity fluctuations responsible for $\psi''_N(0)\to\infty$ in the SSEP can be captured by macroscopic fluctuation theory~\cite{Bertini2015}.   
It is convenient to rescale time by $\tau_L$: let
\beq
\tilde{t} = \frac{t}{\tau_L} = \frac{D_0t}{L^2} \; .
\label{equ:def-tilde-t}
\eeq
Similarly $\tilde\tau = D_0\tau/L^2$.  The space-time thermodynamics approach of Sec.~\ref{sec:spacetime-general} focusses on large deviations with
\beq
-\log p_N(k|\tau) \simeq \tau N {\cal I}(k)
\label{equ:log-p-spacetime}
\eeq
where ${\cal I}(k)$ takes values of order unity.
This is an LDP with speed $\tau$, where the rate function is proportional to $N$.  By contrast, macroscopic fluctuation theory is a theory for large deviations with
\begin{align}
-\log p_N(k|\tau)  & \simeq \tilde{\tau} N \tilde{\cal I}(k) 
\nonumber \\
 & \simeq \tau \tilde{\cal I}(k) \frac{ND_0}{L^2} 
\label{equ:log-p-mft}
\end{align}
with $\tilde{\cal I}(k)$ of order unity.  
From a physical perspective, the interpretation of this formula is that the log-probability of the large deviation is proportional to the system size $N$ and to the time $\tilde{\tau}$, which is measured \emph{on the hydrodynamic scale}.  Just like (\ref{equ:log-p-spacetime}), we interpret (\ref{equ:log-p-mft}) 
as an LDP with speed $\tau$, but now with a rate function proportional to $N/L^2$.  For this one-dimensional system then $L\propto N$ as $N\to\infty$, so the rate function in (\ref{equ:log-p-mft}) goes to zero with system size; this may be  contrasted with (\ref{equ:log-p-spacetime}), where the rate function diverges.  In general, the question of whether (\ref{equ:log-p-spacetime}) or (\ref{equ:log-p-mft}) is applicable depends on whether the fluctuation of interest is governed by hydrodynamic (slow) variables or microscopic (fast) variables.  

For the SSEP,
the macroscopic fluctuation theory gives a quantitative description of fluctuations on the hydrodynamic scale.
They can be analysed by considering a suitable SCGF, for small values of the biasing parameter $s=O(N^{-2})$, see~\cite{Lecomte2012} for details.
The result is that (\ref{equ:log-p-mft}) is applicable for large deviations throughout the range $0<\overline{k}_\tau < \langle k \rangle$.  In this case the fluctuation mechanism is that the SSEP becomes macroscopically inhomogeneous: it forms dense and dilute regions that suppress the activity.  

However, for fluctuations where the intensive activity is significantly larger than $\langle \overline{k}_\tau \rangle$, the probability scales as in (\ref{equ:log-p-spacetime}) and the macroscopic fluctuation theory is not applicable. 
Specifically, for small negative $s$, Ref.~\cite{Appert2008} gives
\beq
{\cal G}(s) \simeq -s\langle k\rangle + A(-s)^{3/2} 
\eeq 
where the constant $A=O(1)$ can be obtained by adapting~\cite[Equ 57]{Appert2008} to the current notation.
From (\ref{equ:calI-legendre}) we see for $k\geq\langle \overline{k}_\tau\rangle$ that
\beq
{\cal I}(k) \simeq  \frac{4}{27A^2} (k-\langle \overline{k}_\tau \rangle)^3 \; .
\eeq
 The second derivative ${\cal G}''(s)$ diverges as $s\to0^-$, while ${\cal I}''(k)$ vanishes as $k\to\langle \overline{k}_\tau\rangle$ from above.  This is consistent with the scaling of the variance of $\overline{k}_\tau$ (inversely proportional to $\tau$ and independent of $N$) and its link to hydrodynamic density fluctuations.
In fact, these fluctuations
have a strong dependence on $s$: for any $s<0$ the system is hyperuniform~\cite{Jack2015}, which means that density fluctuations on large scales are very strongly suppressed~\cite{Torquato2003}.
  
 \subsection{General implications of hydrodynamic modes}
 
We have explained that for the SSEP with periodic boundaries, high-activity fluctuations follow (\ref{equ:log-p-spacetime}) and low-activity fluctuations follow (\ref{equ:log-p-mft}). 
The low-activity regime may be analysed within macroscopic fluctuation theory, which can also be applied to large deviations in other interacting-particle systems, including (weakly) asymmetric exclusion processes and zero-range processes \cite{Bodineau2005,Derrida2007,Bertini2015,Baek2017}.  Similar results can also be found in off-lattice models~\cite{Dolezal2019,Das2019}.

An important general question in this area is whether slow (hydrodynamic) modes lead to fluctuations governed by (\ref{equ:log-p-mft}).
Macroscopic fluctuation theory provides a partial answer.  We use the language of activity fluctuations but the argument is general. We introduce a notion of local equilibration within a spatial region of size $\ell$, with $1\ll \ell \ll L$.  A system is at local equilibrium~\cite{Bertini2015} if the distribution of particles within that region resembles the natural (unbiased) system at the same (local) density.   In this case the hydrodynamic behaviour can be analysed by considering the (smooth) density field, and an associated current.  

 Consider a system in $d$ spatial dimensions so $N\propto L^d$
and suppose that one can construct a macroscopically inhomogeneous state where the (total) activity differs from $\langle k\rangle$ but the system is everywhere in local equilibrium. (Such states may be also be time-dependent, for example travelling waves~\cite{Bodineau2005}, and there may be hydrodynamic flow of particles.)  In this case, macroscopic fluctuation theory explains that the log-probability of fluctuations with this activity obeys (\ref{equ:log-p-mft}).  However, we now have $N\propto L^d$ so the rate function scales as $L^{d-2}$.
The physical interpretation is that local equilibrium states have densities that vary slowly in space: 
these smooth (hydrodynamic) profiles relax slowly towards the steady state and can therefore be stabilised by adding very weak control forces to the system~\cite{Dolezal2019,Das2019}.  This leads to small values of the KL divergence in (\ref{equ:G-var},\ref{equ:Ib-KL}) and hence to small values of the rate function.

In fact, the nomenclature of local equilibrium may be slightly misleading in this context, in that the same argument may be applied to systems with non-equilibrium steady states, such as the active-matter system of Fig.~\ref{fig:active}.  In that case, the same hydrodynamic argument shows that ${\cal I}(\sigma)=0$ for a finite range of $\sigma$ between $0$ and $\langle \sigma \rangle$; see~\cite{Nemoto2019}.  However, this argument relies on the existence of a hydrodynamic theory for this active system where the only relevant field is the density -- general conditions for this to hold in  fluids with non-equilibrium steady states have not yet been established.

\section{Outlook}
\label{sec:outlook}

This article has illustrated some aspects of the rich phenomenology of large deviations of time-averaged quantities.  
The focus has been on the behaviour in large systems, with many interacting degrees of freedom.  In particular, on taking the system size $N\to\infty$, rate functions can develop singular behaviour.  These singularities -- which can be interpreted as dynamical phase transitions -- happen when the mechanism for large fluctuations differs qualitatively from the typical behaviour.  The main examples that we have considered are (i)~the appearance of ferromagnetic order in a $1d$ Ising model~\cite{Jack2010}; (ii)~the existence of an inactive state in the FA model~\cite{Garrahan2007,Bodineau2012cmp,Nemoto2017first}; (iii)~collective motion and arrested phases in an active matter system \cite{Nemoto2019}; (iv)~phase separation and hyperuniformity in the SSEP \cite{Appert2008,Jack2015}.

We emphasise that these examples are illustrative and we have not attempted a comprehensive review.  Among the things that have not been discussed are the recent development of large deviations at level-2.5~\cite{Barato2015,Bertini2018}, which can be interpreted as a more detailed fluctuation theory from which the main results of Sec.~\ref{sec:def} can be derived by the contraction principle, see~\cite{Chetrite2015}.  This theory also allows derivation of thermodynamic uncertainty principles, which are general bounds on the fluctuations of currents, including variances and large deviations~\cite{Gingrich2016,Barato2018-periodic}. In a similar vein, there are some indications that large deviation principles are built on an underlying geometrical structure~\cite{Mielke2014,Maes2008,Kaiser2018}, which has consequences for optimally-controlled processes.

Looking forward, we mention a few directions of ongoing research.  
This review has concentrated on theoretical results and their implications for qualitative behaviour (such as how rate function scale with system size $N$).
However, numerical results have also contributed strongly to large-deviation research.  Building on earlier studies~\cite{Giardina2006,Tailleur2007,Merolle2005,Hedges2009}, recent years have seen renewed interest in efficient and accurate computation of rate functions and SCGFs~\cite{Nemoto2016,Nemoto2017first,Ray2018,Brewer2018,Ray2018jcp,Ferre2018,Jacobson2019,Espigares2019}.  

The theoretical ideas presented here are also being adapted to new settings.  For example, large-deviations of time-averaged quantities are increasingly discussed in \emph{open quantum systems}~\cite{Garrahan2010}, mostly using operator approaches applied to density matrices.  Generalisation of the level-2.5 and optimal control approaches are also being explored in that context~\cite{Carollo2019}.  Another direction of interest is \emph{non-Markovian} models~\cite{Maes2009,Harris2009,Harris2015,Fagg-semi-arxiv}, which can be even richer than the Markovian cases considered here~\cite{Jack2019-growth,Franchini2017}.
Overall, the field has many interesting open questions, and new methods are becoming available, in order to address them.
This makes us optimistic about future progress.

\begin{acknowledgements}
My understanding of large deviation theory and its applications has been shaped by many discussions and collaborations, and I am grateful to many people for their encouragement, advice, and patient explanations.  Special thanks go to Juan P. Garrahan, Peter Sollich, Fred van Wijland, Vivien Lecomte, Hugo Touchette, Freddy Bouchet, Rosemary Harris, Lester Hedges, Mike Evans, Takahiro Nemoto, Paddy Royall, Todd Gingrich, David Limmer, Steve Whitelam, Johannes Zimmer, and Marcus Kaiser.   I would also like to acknowledge fruitful discussions with David Chandler, whose advice and enthusiastic support were extremely important to me.
\end{acknowledgements}

\begin{appendix}

\section{Optimal control calculation}
\label{app:control}

\subsection{Equivalence of optimal-control problem and eigenproblems for a stochastic differential equation}

For large deviations of $\overline{b}_\tau$ in the model of (\ref{equ:sde-simple}), we show that using the controlled model (\ref{equ:sde-con}) with (\ref{equ:psi-var}) and maximising over $\phi$ is equivalent to solving the eigenproblem (\ref{equ:WW-simple}). 
Using the theory of path integrals with Stratonovich convention,
we write the path probability\footnote{%
The integral $\int_0^\tau |\partial_t \CC_t |^2\mathrm{d}t$ that appears in (\ref{sec:path-int}) is not mathematically well-defined for processes like (\ref{equ:sde-simple}). This may be resolved by defining $P_\tau(\bm{\CC})$ as a density with respect to the path-measure for a Brownian motion (see for example~\cite{Kaiser2017ent}), or by considering $P(\bm{\CC})$ to be the probability density of an explicitly time-discretised trajectory.  In either case one arrives at the same result in (\ref{equ:path-int-rel}), which is well-defined and unambiguous.
} 
for (\ref{equ:sde-simple}) as
\beq
P_\tau(\bm{\CC}) \propto \pi(\CC_0) \exp\left(  - \int_0^\tau \frac{\left|\partial_t \CC_t - v(\CC_t)\right|^2}{4} + \frac{\nabla\cdot v(\CC_t)}{2} \, \mathrm{d}t \right) \; .
\label{sec:path-int}
\eeq 
where $\pi$ is the probability of the initial condition.
A similar expression holds for the controlled process (\ref{equ:sde-con}): we assume for simplicity that this process has the same initial distribution $\pi$ as the original process, although this assumption is easily relaxed.
In order to apply (\ref{equ:psi-var}) we compute
\begin{multline}
\log \frac{ P^{\rm con}_\tau(\bm{\CC}) }{ P_\tau(\bm{\CC}) } =  
-\frac12 \int_0^\tau \nabla\phi(\CC_t) \circ \mathrm{d}\CC_t
\\
+\frac12 \int_0^\tau  \nabla\phi(\CC_t) \cdot v(\CC_t) + \nabla^2\phi(\CC_t) - \frac12|\nabla\phi(\CC_t)|^2 \mathrm{d}t   
\label{equ:path-int-rel}
\end{multline}
where the $\circ$ indicates a Stratonovich product.  This gives an explicit expression for $\overline{g}_\tau$ in (\ref{equ:rk-g}).
The integral in the first line can be evaluated as $[\phi(\CC_\tau)-\phi(\CC_0)]$.

To apply (\ref{equ:psi-var}) we require the average of (\ref{equ:path-int-rel}), as $\tau\to\infty$.  Ergodicity of the controlled process allows us to replace averages of  time integrals by averages with respect to the steady-state distribution, which we denote by $\mu$.  So (\ref{equ:psi-var}) becomes
\beq
\psi(s) \geq \frac12 \int  \left[ - 2 s b - \nabla\phi \cdot v - \nabla^2\phi + \frac12|\nabla\phi|^2  \right] \mu\, \mathrm{d}\CC
\label{equ:psi-var-sde}
\eeq
It is not necessary to compute $\mu$ explicitly, one uses instead the Fokker-Planck equation for the controlled process to show that it solves 
\beq
\nabla\cdot[v\mu-\mu\nabla\phi] = \nabla^2\mu \; 
\label{equ:fp-con}
\eeq
and one also has $\int \mu(\CC) \mathrm{d}\CC=1$.  These are two constraints that can be implemented by  Lagrange multipliers: we are left to find an extremum of
\begin{multline}
\frac12 \int  \left[ - 2 s b -\nabla\phi \cdot v - \nabla^2\phi + \frac12|\nabla\phi|^2  \right] \mu \mathrm{d}\CC
\\ + \frac12 \int \left[  \nabla\lambda  \cdot ( v\mu-\mu\nabla\phi-\nabla\mu ) + 2\gamma \mu \right] \mathrm{d}\CC 
\end{multline}
where the functional Lagrange multiplier $\lambda$ enforces (\ref{equ:fp-con}) while $\gamma$ enforces normalisation of $\mu$.  
A short calculation shows that the extremum occurs for $\lambda=0$, and is characterised by
\beq
-\nabla\phi \cdot v - \nabla^2\phi + \frac12|\nabla\phi|^2 - 2 s b = 2\gamma
\label{equ:HJ}
\eeq
This is an example of a Hamilton-Jacobi equation (or a Hamilton-Jacobi-Bellman equation).
Using it with (\ref{equ:psi-var-sde}) shows that solutions of the variational problem have $\gamma\leq\psi(s)$.
Moreover, writing $\phi = -2\log {\cal F}$ yields
\begin{align} 
\gamma {\cal F} 
& = \nabla^2 {\cal F} + v \cdot \nabla {\cal F} - s b {\cal F}
\nonumber\\
& = {\cal W}_s^\dag {\cal F} 
\label{equ:F-W}
\end{align}
where the second line follows from the expression for ${\cal W}_s$ given in (\ref{equ:WW-simple}).  This is an eigenfunction equation for the operator ${\cal W}_s^\dag$, and $\gamma$ is the associated eigenvalue.  

The optimal bound on $\psi$ is obtained by taking the largest available solution for $\gamma$, which is therefore the largest eigenvalue of ${\cal W}_s^\dag$ -- this
is equal to $\psi(s)$, by (\ref{equ:WW-simple}).
It follows that (\ref{equ:psi-var-sde}) is an equality if one takes the (optimal) control potential $\phi=-2\log{\cal F}$ where ${\cal F}$ is the relevant eigenfunction of ${\cal W}_s^\dag$.

\subsection{Example: large deviations of squared displacement in an Ornstein-Uhlenbeck process}

To illustrate this general discussion, we analyse the specific case of a one-dimensional Ornstein-Uhlenbeck process, which  is
\beq
\mathrm{d}x_t = - \omega x_t \mathrm{d}t + \sqrt{2}\mathrm{d}W_t
\label{equ:dx-OU}
\eeq
where $x_t$ is a real number.  The force $-\omega x$ is the gradient of a potential $\frac12 \omega x^2$.  
Large deviations for this process have been discussed previously in several contexts, for example~\cite{Majumdar2002,Touchette2018-lectures,Nickelsen2018}.
We consider large deviations of $\overline{b}_\tau = \tau^{-1}\int_0^\tau x_t^2 \mathrm{d}t$ so $b(x_t)=x_t^2$.
The tilted generator of (\ref{equ:WW-simple}) is ${\cal W}_s$, which acts on probability density functions $P$ as
\beq
{\cal W}_s P = \nabla \cdot ( \omega x P + \nabla P ) - sx^2 P  \; .
\eeq
Its largest eigenvalue is the SCGF
\beq
\psi(s) = (\omega/2) - \sqrt{ s + (\omega^2/4) } \; .
\label{equ:psi-OU}
\eeq
This result is valid for $s>-\omega^2/4$, otherwise the spectrum of ${\cal W}_s$ is not bounded from below, this is linked to the behaviour of the rate function, as we discuss below.
The associated eigenfunction is 
\beq
P_{\rm end}(x|s) = \sqrt{\frac{\omega-\psi(s)}{2\pi}} \exp\left[ -\frac{x^2(\omega-\psi(s))}{2} \right] \; .
\eeq
Note that $\psi(s)\leq(\omega/2)$ so this is a normalised probability density function, for $\psi(s)=0$ it reduces to the steady-state distribution of (\ref{equ:dx-OU}).
Using (\ref{equ:psi-OU}) with (\ref{equ:I-legendre}), the rate function for $\overline{b}_\tau$ is
\beq
I(b) = \frac{(  b\omega - 1)^2}{4b} \; .
\label{equ:I-OU}
\eeq
For the original OU process one has $\langle b \rangle = (1/\omega)$, so $I(\langle b \rangle)=0$, as required.
The fact that the SCGF only exists for $s>-\omega^2/4$ reflects the fact that the rate function grows linearly as $b\to\infty$, where it scales as $I(b) \sim b\omega^2/4$.  Hence its Legendre transform $\psi(s)$ only exists if $s$ is sufficiently large.
The solution of the backward Fokker-Planck equation (\ref{equ:F-W}) for this process is
\beq
{\cal F}(x|s) = \exp\left( \frac{x^2 \psi(s)}{2} \right) \; .
\eeq
From this we infer (using $\phi=-2\log{\cal F}$) that the optimal control potential $\phi = -\psi(s) x^2$.  For $s>0$ then $\psi(s)<0$ and the control potential results in an additional confining force that reduces the typical value of $x_t^2$.

To verify that these results are consistent with the optimal-control theory, recall that the controlled process in this case is
\beq
\mathrm{d}x_t = - (\omega x_t + \nabla\phi(x_t))\mathrm{d}t + \sqrt{2}\mathrm{d}W_t \; .
\eeq
Identifying $v=-\omega x$ then (\ref{equ:HJ}) becomes
\beq
\omega x \nabla\phi  - \nabla^2\phi + \frac12|\nabla\phi|^2 - 2 s x^2  = 2\gamma \; .
\label{equ:HJ-OU}
\eeq
This equation can be solved by taking $\gamma = \psi(s)$ [as given in (\ref{equ:psi-OU})] and $\phi = -\psi(s)x^2$.
This is indeed the optimal control potential $\phi$  that maximises the bound in (\ref{equ:psi-var-sde}), which then becomes an equality.

To understand this result in a more intuitive way, it is useful to restrict to a controlled process with a quadratic control potential $\phi(x) = \frac12 ux^2$, where $u$ is a variational parameter.
Then (\ref{equ:path-int-rel}) becomes
\begin{equation}
\log \frac{ P^{\rm con}_\tau(\bm{\CC}) }{ P_\tau(\bm{\CC}) } =  
\frac{u}{4} [ x_0^2 - x_\tau^2 ]
+\frac12 \int_0^\tau  \left[ -\omega u x_t^2 +u - \frac{u^2 x_t^2}{2} \right] \mathrm{d}t   
\end{equation}
The controlled process describes motion in a potential $\frac12(\omega+u)x^2$ so $\langle x_t^2 \rangle_{\rm con} = 1/(\omega+u)$ and one obtains by (\ref{equ:KL})
\beq
 \lim_{\tau\to\infty} \frac{1}{\tau} {\cal D}(P^{\rm con}_\tau || P_\tau) 
 =\frac{u^2}{4(\omega+u)}
 \label{equ:KL-OU}
\eeq
This quantity measures how different is the controlled process from the original OU process.  For $u=0$ the two processes coincide and the KL divergence is zero.

Moreover, this controlled process has $\langle b\rangle_{\rm con}=1/(\omega+u)$ so it can be used with (\ref{equ:Ib-KL}), as long as one takes $b=\langle b\rangle_{\rm con}$, that is $u=b^{-1}-\omega$.  The  result is a bound on the rate function
\beq
I(b) \leq \frac{b}{4}\left( \omega - b^{-1} \right)^2 \; .
\label{equ:rate-OU-v2}
\eeq
Comparison with (\ref{equ:I-OU}) shows that this variational result holds as an equality.  This occurs because the ansatz of a quadratic control potential is sufficient to capture the optimal control.  
It follows that the optimally-controlled dynamics for fluctuations with $\overline{b}_\tau=b$ is simply 
\beq
\mathrm{d}x_t = - (x_t/b) \mathrm{d}t + \sqrt{2}\mathrm{d}W_t \;.
\eeq
The SCGF can also be obtained by using (\ref{equ:KL-OU}) with (\ref{equ:psi-var}), or equivalently by using (\ref{equ:rate-OU-v2}) with (\ref{equ:psi-legendre}).

The conclusion of this analysis for the OU process (\ref{equ:dx-OU}) is that large deviations of the time-average of $x_t^2$ occur by trajectories that are representative of a similar (controlled) OU process, in which only the parameter $\omega$ is modified.  This parameter governs the size of the restoring force in (\ref{equ:dx-OU}) and hence the typical value of $\overline{b}_\tau$.

\section{Glauber-Ising model}
\label{app:ising}

This section summarises some results for large deviations of the time-integrated energy in the one-dimensional Glauber-Ising model. 
This situation was analysed in~\cite{Jack2010}.  Here we summarise the results and we also correct two small errors in that analysis.  
Details of the corrected analysis are given in~\cite{Guioth-arxiv}.

\newcommand{\sig}{\sigma}
The Glauber-Ising chain has $N$ spins $\sig_i=\pm 1$.  The energy is $E=-\frac12 \sum_i \sig_i \sig_{i+1}$ with periodic boundaries.  Spin $i$ flips with rate 
$1/(1+\ee^{\beta h_i \sig_i})$ where $\beta$ is the inverse temperature, and $h_i=(\sig_{i-1}+\sig_{i+1})$ is the local field on site $i$, such that $h_i \sig_i$ is the change in energy on flipping spin $i$.
The model can be analysed by writing $\tau_i=\sig_{i}\sig_{i+1}$ so that $\tau_i=-1$ if there is a domain wall between spin $i$ and spin $i+1$, and $\tau_i=+1$ otherwise. 
Then the energy is $-\sum_i \tau_i/2$.  The domain walls can be interpreted as particles that evolve according to a reaction-diffusion dynamics.   Since the system has periodic boundaries then it is important to note that the number of these domain walls is always even.

In the domain-wall representation, the operator ${\cal W}_s$ can be represented in terms of Pauli matrices.
The dependence of the model on temperature is incorporated through a parameter $\lambda=2/(1+\ee^{2\beta})\leq 1$.    
As stated in~\cite{Jack2015}, the operator ${\cal W}_s$
can be diagonalised using a Jordan-Wigner transformation.  
 Some details of a similar computation are given in~\cite{Grynberg1994}, where it is emphasised
that the Jordan-Wigner transformation requires some care with the periodic boundary conditions.  Using that the number of domain walls is always even the relevant operator can be diagonalised in a Fourier basis as
\beq
{\cal W}_s = \frac12 \sum_q \left[ \Omega_q ( \beta_q^\dag \beta_q - \beta_q \beta_q^\dag ) -1 \right]
\label{equ:W-ff}	
\eeq
where the sum runs over wavevectors $q$ in the first Brillouin zone (see below), $\beta_q$ and $\beta_q^\dag$ are fermionic creation and annihilation operators, and
\beq
\Omega_q = \sqrt{ (1-\cos q+s-\lambda)^2 + \lambda(2-\lambda) \sin^2 q }
\eeq
Compared with~\cite{Jack2010}, we have corrected a factor of two in (\ref{equ:W-ff}).  In addition, careful treatment of the interplay between periodic boundaries and the Jordan-Wigner transformation requires that the wavevectors $q$ in (\ref{equ:W-ff}) are $q=(2m+1)\pi/N$ with $m=0,1,2,\dots,L-1$ so that $qN/\pi$ is an odd integer~\cite{Grynberg1994}, contrary to~\cite{Jack2010}.  Deriving this result uses explicitly that the number of domain walls is even.  
The largest eigenvalue of ${\cal W}_s$ is then
\beq
\psi_N(s) =  \frac12 \sum_q \left( \Omega_q - 1 \right)
\eeq
The corresponding eigenvector $|0\rangle$ obeys $\beta_q^\dag|0\rangle=0$ for all $q$.   For $s=0$ it can be checked that $\Omega_q=[1+(1-\lambda)\cos q]$ so $\psi(0)=0$, as required.   

The function $\psi_N(s)$ is analytic and convex in $s$, as it must be because the model is in the class of Sec.~\ref{sec:scope}.
Taking the large-$N$ limit, the free energy per site of (\ref{equ:GG-lim}) is
\beq
{\cal G}(s) =  \frac12 \int_{-\pi}^\pi \left( \Omega_q - 1 \right) \frac{\mathrm{d}q}{2\pi} \; .
\eeq
This function has a singularity in its second derivative at the phase transition point $s=\lambda$, at which $\Omega_{q}\to0$ for small-$q$.  This is a dynamical phase transition model and the system is ferromagnetic for $s>\lambda$.

\end{appendix}

\bibliographystyle{apsrev4-1}       
\bibliography{dev}   

\end{document}